%% file: main.tex
\documentclass{article}

\PassOptionsToPackage{numbers,compress}{natbib}
\usepackage[main,final]{neurips_2026}

\usepackage[utf8]{inputenc}
\usepackage[T1]{fontenc}
\usepackage{hyperref}
\usepackage{url}
\usepackage{booktabs}
\usepackage{amsfonts}
\usepackage{amsmath}
\usepackage{amssymb}
\usepackage{nicefrac}
\usepackage{microtype}
\usepackage{xcolor}
\usepackage{graphicx}
\usepackage{subcaption}

\newif\ifshowchanges
\showchangesfalse
\hypersetup{hidelinks}
\definecolor{revisionred}{RGB}{185,20,35}
\newcommand{\rev}[1]{{\ifshowchanges\color{revisionred}\fi #1}}
\newenvironment{revision}{\begingroup\ifshowchanges\color{revisionred}\captionsetup{font={color=revisionred}}\fi}{\par\endgroup}
\title{Spike-to-Field Mechanisms of Turbulence-Like Dynamics in Spatial Spiking Neural Networks}

\author{%
  \rev{Zequan Liu}\\
  \rev{ZeQome}\\
  \rev{Zurich, Switzerland}\\
  \rev{\texttt{zequan@zeqome.com}}
}

\begin{document}

\maketitle

\begin{abstract}
Neural avalanches and turbulence-like brain dynamics are usually measured after neural activity has been coarse-grained into continuous fields, leaving open which spike-level mechanisms can generate these macroscopic observables. We build a spatial leaky integrate-and-fire spiking neural network and explicitly transform spikes into activity-density and Hilbert phase fields. A delayed, distance-dependent, passive-dendrite regime reproducibly generates turbulence-like field structure, including high-amplitude phase-defect tracks, structure-function scaling, and power-law-favored continuous-field avalanche tails. Controls show that firing rate, vortex count, and avalanche tail fits are individually insufficient: destroying delayed propagation, dendritic filtering, or spike timing suppresses key field signatures, while random-space nulls can retain many mathematical phase defects \rev{but show weaker spatial structure and synaptic-delay transfer}. Biologically motivated gates further shape the regime: PV-like perisomatic inhibition clamps phase dynamics, SST-like dendritic inhibition and projection-specific inhibitory facilitation preserve long-lived tracks, and a rate-constrained projection-STP regime \rev{expresses a roughly 200-ms episode at 12.25 Hz (full-record mean)} across 12 seeds. These results use SNNs as a mechanistic testbed for spike-derived brain turbulence observables and show why phase defects and avalanche-like statistics must be interpreted jointly with spatial scaling and spike-level propagation.
\end{abstract}

\section{Introduction}

Large-scale brain activity is often modeled as a continuous dynamical field, even though it ultimately emerges from discrete spikes and synaptic events. This separation of levels is useful, but it leaves a mechanistic gap: if turbulence-like phase and amplitude statistics are measured in empirical neuroimaging, what spike-level circuit mechanisms can generate them? Conversely, if neuronal avalanches are measured as scale-free cascades in activity, can they be connected to the same spatial phase structures used in turbulence analyses?

Recent work on human brain dynamics has proposed that intrinsic activity can exhibit turbulence-like signatures that support cross-scale information transfer \citep{deco2020turbulent}. Related work on cortical travelling and metastable waves shows that neural activity can organize into propagating, rotating, and source/sink-like spatiotemporal patterns across scales \citep{muller2018travelling,roberts2019metastable}. In parallel, the neuronal avalanche literature has emphasized scale-free activity cascades as a signature of critical dynamics \citep{beggs2003neuronal,friedman2012universal,fontenele2019criticality}. Both lines of work rely on a coarse-grained description: empirical fMRI, MEG, LFP, or binned population activity is treated as a field. The present paper studies the spike-level mechanisms that could support such turbulence dynamics by asking whether a minimal spatial SNN can generate the same classes of observables after an explicit spike-to-field transformation.

We focus on mechanism rather than whole-brain reconstruction. A small two-dimensional SNN is not intended to replace empirical studies of human brain turbulence; instead, it gives an executable bridge between three descriptions: binary spike rasters, continuous activity-density and phase fields, and turbulence-like statistics. This bridge allows us to test which microscopic ingredients \rev{support measured field organization}: distance-dependent recurrent connectivity, finite conduction delays, excitation/inhibition balance, dendritic filtering, and projection-specific short-term synaptic plasticity (STP).

We use the term ``turbulence-like'' operationally. In this paper it denotes a joint pattern of spike-derived field observables: spatially structured activity-density fluctuations, phase-defect tracks in high-amplitude regions, structure-function scaling, and spike-level delayed propagation. It does not mean that the model solves the Navier--Stokes equations, reproduces a whole human brain, or proves that any single graph motif is sufficient for empirical brain turbulence. \rev{These observables are not sufficient evidence of physical turbulence.}

The contributions are:
\begin{itemize}
    \item A reproducible pipeline from discrete SNN spikes to continuous density fields, Hilbert phase fields, phase vortices, avalanche candidates, and spatial structure functions.
    \item A mechanism-guided passive-dendrite delayed-propagation regime with reproducible high-amplitude phase vortices across 12 random seeds.
    \item Control experiments showing that avalanche tail fits alone are insufficient: vortex density and vortex track lifetime distinguish mechanisms more sharply than power-law-vs-exponential avalanche comparisons.
    \item A separation between localized turbulence-field assays and classical avalanche assays: sparse post-stimulus rasters do not support a Beggs--Plenz criticality claim, whereas long weak-drive spontaneous-like records produce many spike-bin avalanches with branching ratios near one.
    \item A biologically motivated extension in which dendritic inhibition and projection-specific STP modulate turbulence-like dynamics, with SST-like dendrite-targeting inhibitory facilitation producing the \rev{highest long-track count} in the tested set.
    \item A topology control showing that feedforward graphs favor direct pairwise spike transfer, whereas recurrent graphs preserve rate-matched activity re-entry, separating transmission from \rev{activity re-entry.}
\end{itemize}

\section{Related work}

\paragraph{Brain turbulence and neural fields.}
Turbulence-like brain dynamics have been studied in whole-brain empirical and oscillator models, where amplitude fluctuations and their spatial correlations are used to quantify cross-scale communication \citep{deco2020turbulent}. These approaches are related to a longer tradition of neural field and large-scale dynamical modeling \citep{wilson1972excitatory,breakspear2017dynamic}, cortical travelling waves \citep{muller2018travelling}, metastable brain waves \citep{roberts2019metastable}, and mesoscopic phase-vortex descriptions of brain activity \citep{freeman2009vortices}. Our goal is complementary: we study the circuit-level mechanisms that can generate the continuous-field observables used to study brain turbulence, starting from spikes in a spatially embedded recurrent circuit.

\paragraph{Neuronal avalanches.}
Neuronal avalanches were introduced as scale-free cascades in cortical circuits \citep{beggs2003neuronal}. Later work has connected avalanche statistics to critical dynamics, scaling functions, cortical state transitions, and possible computational benefits \citep{friedman2012universal,fontenele2019criticality,shew2013functional}. Standard avalanche analyses use binned events and cluster sizes; our main turbulence-field pipeline computes connected high-activity volumes in a continuous density field. This is closer to a non-equilibrium field-theoretic view of avalanches, but it inherits threshold and smoothing choices. Because power-law-like tails can arise spuriously and sparse sampling can distort inferred branching statistics \citep{clauset2009power,touboul2010powerlaw,priesemann2009subsampling,wilting2018inferring}, we also include a direct traditional spike-bin diagnostic on sparse electrode samples, distinguishing short post-stimulus records from long weak-drive spontaneous-like records.

\paragraph{Dendrites, inhibition, and STP.}
Dendrites are not passive cables in all regimes: pyramidal neurons contain domain-specific synaptic inputs, active conductances, and dendritically initiated events \citep{spruston2008pyramidal,major2013active}. Inhibition is also strongly target- and cell-type-specific, with PV-like perisomatic and SST-like dendritic effects representing different network controls. Short-term synaptic plasticity can be projection-specific: a single axon can signal differently to different targets \citep{markram1998differential}, and inhibitory microcircuits use diverse synaptic dynamics \citep{gupta2000organizing}. These facts motivate our simplified tests of passive dendrites, active dendrites, dendritic inhibition, and projection-specific STP.

\section{Model and analysis pipeline}

\subsection{Spatial LIF network}

We simulate a $24 \times 24$ grid of leaky integrate-and-fire neurons. Each neuron is excitatory with probability 0.8 and inhibitory otherwise. The membrane voltage obeys
\begin{equation}
    \tau_m \frac{dv_i}{dt}
    = E_L - v_i + I_0 + I_i^{\mathrm{stim}}(t) + I_i^{\mathrm{syn}}(t) + \xi_i(t),
\end{equation}
with reset and refractory dynamics after threshold crossing. A short stimulus is applied to a fixed left-side patch only to initiate activity; subsequent waves and activity clusters are generated by recurrent interactions.

Outgoing connections are sampled according to an exponential distance rule,
\begin{equation}
    p_{ij} \propto \exp(-\|r_i-r_j\|/\lambda),
\end{equation}
with separate out-degree parameters for excitatory and inhibitory neurons. Conduction delays are distance-dependent:
\begin{equation}
    d_{ij}=1+\mathrm{round}(\|r_i-r_j\|/v_c),
\end{equation}
clipped by a maximum delay. This makes wavefronts and collisions possible without prescribing a moving wave in the input, consistent with the broader view that recurrent cortical circuits can support travelling-wave dynamics \citep{muller2018travelling}.

The passive-dendrite candidate used in the main experiments has grid size 24, 340 time steps, passive dendrites, $\lambda=8.55$ cells, conduction speed $v_c=1.06$ cells/ms, $k_E=k_I=32$, excitatory weight 11.97, inhibitory weight magnitude 12.31, tonic drive 11.31 mV, noise standard deviation 1.36 mV, stimulus amplitude 26.45 mV, dendrite coupling 0.759, and dendrite time constant 20.19 ms.

\subsection{Spike-to-field transformation}

Let $s_i(t)\in\{0,1\}$ denote whether neuron $i$ spikes at time $t$. We reshape spikes onto the two-dimensional grid and apply separable Gaussian smoothing in time and space:
\begin{equation}
    \rho(r,t) = \frac{1}{\Delta t}
    \sum_{i,t'} s_i(t') G_{\sigma_r}(r-r_i) G_{\sigma_t}(t-t').
\end{equation}
The resulting $\rho(r,t)$ is a continuous activity-density field in Hz. To obtain a phase field, we band-pass filter $\rho$ in the 20--80 Hz range and apply the Hilbert transform at each spatial location:
\begin{equation}
    z(r,t)=A(r,t)e^{i\theta(r,t)}.
\end{equation}
Here $A(r,t)$ is the \rev{local band-limited amplitude} and $\theta(r,t)$ is the phase field. \rev{Gaussian widths are 2 ms/1.2 cells; filtering uses third-order forward--backward Butterworth.}

\subsection{Turbulence-like observables}

We compute phase vortices as topological defects on each plaquette. For plaquette $p$ with cyclic corner phases $\theta_{p,1},\ldots,\theta_{p,4}$ and $\theta_{p,5}=\theta_{p,1}$,
\begin{equation}
    q_p = \frac{1}{2\pi}\sum_{k=1}^{4} \mathrm{wrap}(\theta_{p,k+1}-\theta_{p,k}),
\end{equation}
where $q_p=\pm 1$ indicates a candidate vortex. We count only defects whose surrounding amplitude exceeds \rev{the 60th percentile}, because phase is unreliable in low-amplitude regions. Vortices are linked across frames by same-sign nearest-neighbor matching with \rev{a maximum distance of 2.6 cells and gap of two frames}, yielding track lifetimes.

We estimate continuous-field avalanche candidates by thresholding $\rho$ and computing three-dimensional connected components in $(x,y,t)$. We fit discrete power-law tails by maximum likelihood over candidate $x_{\min}$ values and compare against an exponential tail using AIC, following the principle that power-law claims require likelihood-based tail fitting rather than visual log-log regression \citep{clauset2009power}. We also compute a second-order spatial structure function,
\begin{equation}
    S_2(\ell)=\left\langle |\rho(r+\ell,t)-\rho(r,t)|^2 \right\rangle_{r,t},
\end{equation}
and report the log-log slope over the central lag range.

The key point is that no single observable defines turbulence-like dynamics in this setting. Density fields measure amplitude organization, phase fields measure local oscillatory timing, vortices measure phase topology, structure functions measure spatial scaling, and synaptic-delay transfer checks whether field structure is supported by spike-level propagation.

\section{Experiments}

The experiments are organized around a compact main evidence chain. First, we show that a delayed spatial SNN can reproducibly generate spike-derived field structure. Second, we show that avalanche fits and vortex counts alone are insufficient by using mechanistic and rate-matched controls. Third, we test biological gates--dendritic inhibition, PV/SST/VIP microcircuits, and projection-specific STP--that move the same field dynamics into a lower-rate regime. Fourth, we separate two avalanche settings: sparse post-stimulus rasters are not enough to claim classical criticality, while long weak-drive spontaneous-like records can produce traditional spike-bin avalanches with branching ratios near one. Finally, a topology control separates direct spike transfer from recurrent field re-entry. Additional diagnostic figures are provided in Appendix~\ref{app:additional_diagnostics}.

\subsection{Candidate regime and reproducibility}

We first instantiated a mechanism-guided passive-dendrite operating point combining finite delayed propagation, balanced excitation/inhibition, noise-driven variability, and a localized trigger. The pilot candidate produced mean firing rate 105.18 Hz, mean high-amplitude vortices 6.58 per sampled frame, peak high-amplitude vortices 37, vortex occupancy 0.54, and $S_2$ slope 0.965.

We then fixed the candidate parameters and reran 12 random seeds, resampling network connectivity, initial voltages, and noise. All 12 seeds passed \rev{an exploratory field criterion}. The mean firing rate was 115.43 Hz, mean high-amplitude vortices were 6.05, mean vortex tracks with lifetime at least 8 frames were 82.5, and the mean structure-function slope was 0.94.

This baseline should be interpreted as a high-excitability operating point rather than a biologically typical spontaneous cortical firing-rate regime. Its role is to expose a reproducible spike-derived field state before adding circuit mechanisms that can regulate activity level. The projection-specific STP experiment below shows that similar long-lived phase dynamics \rev{can also occur episodically} at much lower mean firing rate.

\subsection{Avalanche tail fitting}

Pooling avalanche candidates across the 12-seed candidate regime yielded 1825 samples. The fitted power-law tail had $x_{\min}=4$, 747 tail samples, exponent $\alpha=2.06$, and $\Delta\mathrm{AIC}=\mathrm{AIC}_{\exp}-\mathrm{AIC}_{\mathrm{power}}=5347.2$, favoring the power-law tail over the exponential alternative. Each of the 12 individual seeds also favored the power-law tail by AIC.

This result is encouraging but not by itself sufficient evidence for turbulence. Avalanche statistics depend on thresholding, smoothing, and finite-size effects, and several controls below also show power-law-favored tails. We therefore treat avalanche tail fits as one axis of evidence, not the defining criterion. \rev{Five-family tail comparisons retain power-law preference in 12/12 high-rate records, versus 8/11 episodes; cutoffs remain power-law-selected.}

\subsection{Mechanistic controls}

Table~\ref{tab:controls} summarizes the main controls. Removing distance-dependent delays reduced mean high-amplitude vortices from 6.05 to 3.70 and long tracks from 82.5 to 58.3. Removing passive dendrites reduced vortices further to 2.29 and long tracks to 37.25 while increasing mean firing rate to 164.25 Hz. A short-range delayed regime preserved vortex statistics, suggesting that finite delayed propagation is more important than simply having long connections.

\begin{table}[t]
    \centering
    \caption{Mechanistic controls over 12 seeds. Power-law-favored avalanche tails remain common, while vortex statistics distinguish propagation and dendritic mechanisms more sharply.}
    \label{tab:controls}
    \begin{tabular}{lrrrrr}
        \toprule
        Model & Rate & Vortices & Tracks $\geq 8$ & $S_2$ slope & $\alpha$ \\
        \midrule
        Passive + distance delay & 115.43 & 6.05 & 82.50 & 0.94 & 2.05 \\
        No distance delay & 128.67 & 3.70 & 58.33 & 0.96 & 2.01 \\
        Short range + delay & 113.65 & 6.10 & 92.00 & 1.15 & 1.77 \\
        No passive dendrite & 164.25 & 2.29 & 37.25 & 0.98 & 2.06 \\
        Uniform fast control & 129.23 & 4.28 & 64.83 & 0.84 & 1.92 \\
        \bottomrule
    \end{tabular}
\end{table}

\subsection{\rev{Baseline and mechanism-control field morphology}}

To make the contrast explicit, we consolidated the candidate regime and several \rev{mechanism-ablated or inhibition-clamped controls} into one comparison. \rev{The baseline candidate} has 6.05 mean high-amplitude vortices and 82.5 long tracks. Removing distance-dependent delays reduces these to 3.70 vortices and 58.33 long tracks. Removing passive dendrites increases firing rate to 164.25 Hz but reduces vortices to 2.29 and long tracks to 37.25, showing that high firing rate alone is not sufficient for turbulence-like dynamics. A PV-clamped microcircuit is an even stronger negative control, with only 0.35 vortices and 4.08 long tracks. In contrast, an SST-only dendritic-inhibition circuit preserves turbulence-like dynamics, with 5.93 vortices and 94.33 long tracks.

\begin{figure}[t]
    \centering
    \includegraphics[width=0.96\linewidth]{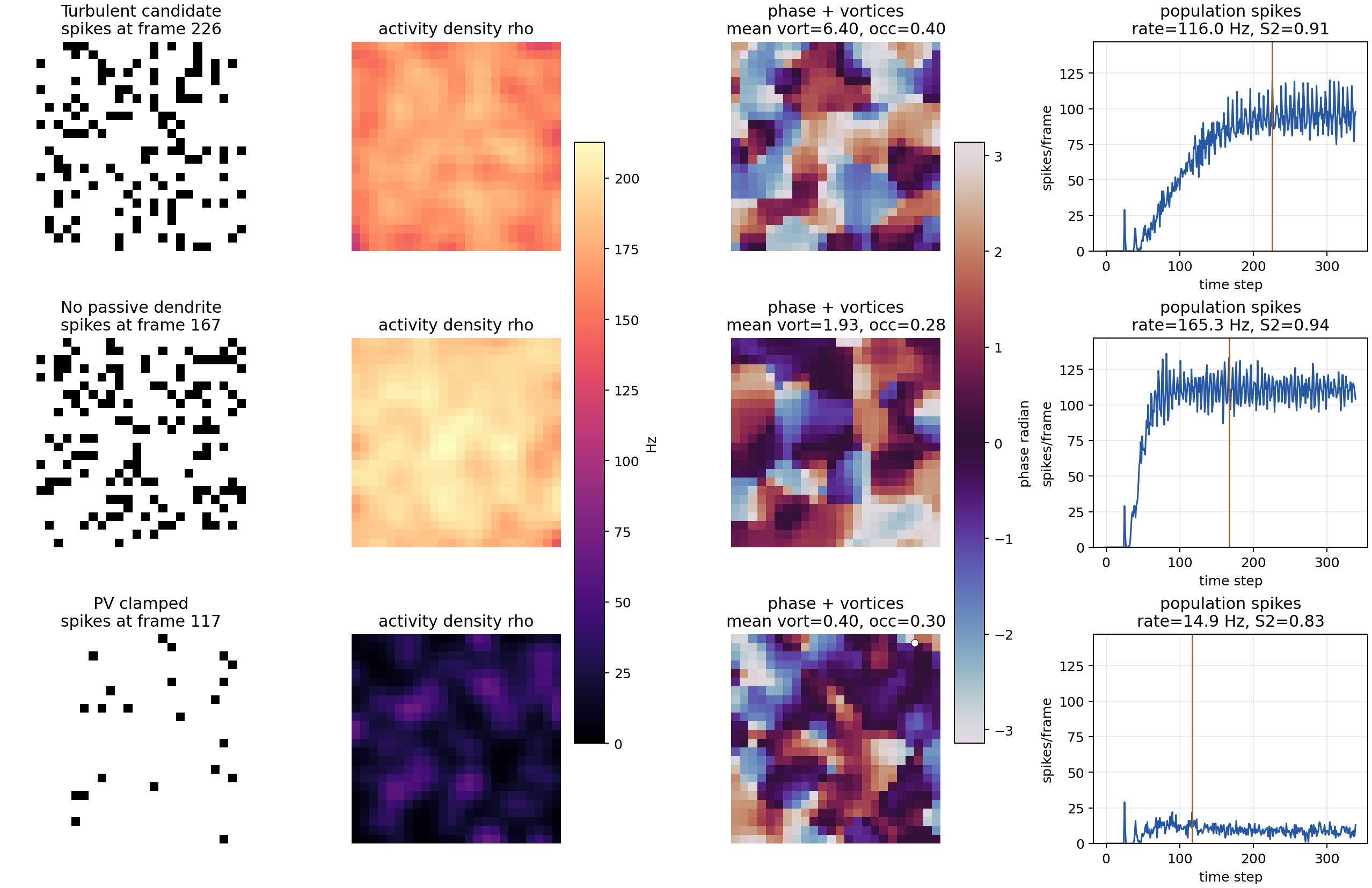}
    \caption{Visual spike-to-field comparison. Each row is a different SNN regime and each column shows a different representation of the same simulated activity: raw spikes on the grid, smoothed activity density $\rho$, Hilbert phase with high-amplitude vortex markers, and the population spike-count trace. The turbulence-like candidate has structured spike clusters that coarse-grain into moving density and phase patterns; the no-passive-dendrite high-rate control fires strongly but produces weaker long-lived phase organization; the PV-clamped control suppresses both spiking and field structure.}
    \label{fig:turbulent_vs_non_snapshots}
\end{figure}

Figure~\ref{fig:turbulent_vs_non_snapshots} gives the visual version of this comparison. The left column shows the raw spiking raster on the two-dimensional grid before any field analysis. The next columns show how those same spikes become a smoothed activity-density field, a phase field with detected high-amplitude vortices, and a population spike-count trace. This figure is important because the field observables are not introduced as abstract post-processing objects; they are visually traceable back to discrete spike patterns.

\begin{figure}[t]
    \centering
    \begin{subfigure}{0.98\linewidth}
        \centering
        \includegraphics[width=\linewidth]{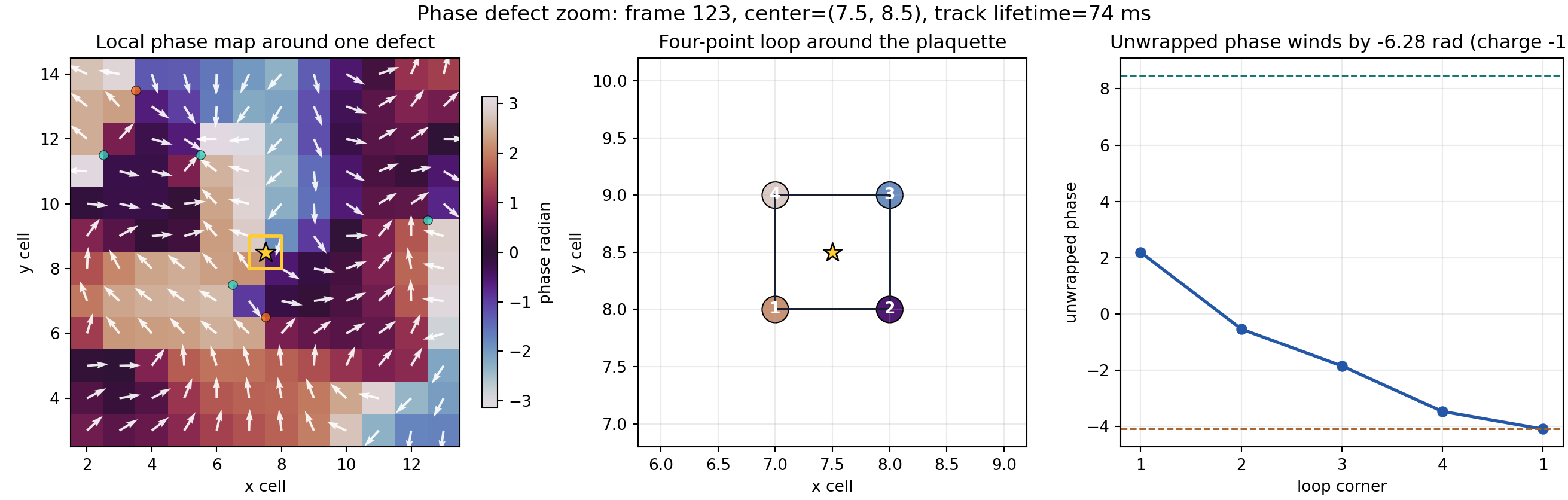}
        \caption{Local phase-defect zoom. The highlighted plaquette has a wrapped phase winding of approximately $-2\pi$, making the center a topological phase defect rather than an ordinary high-activity point.}
    \end{subfigure}
    \begin{subfigure}{0.82\linewidth}
        \centering
        \includegraphics[width=\linewidth]{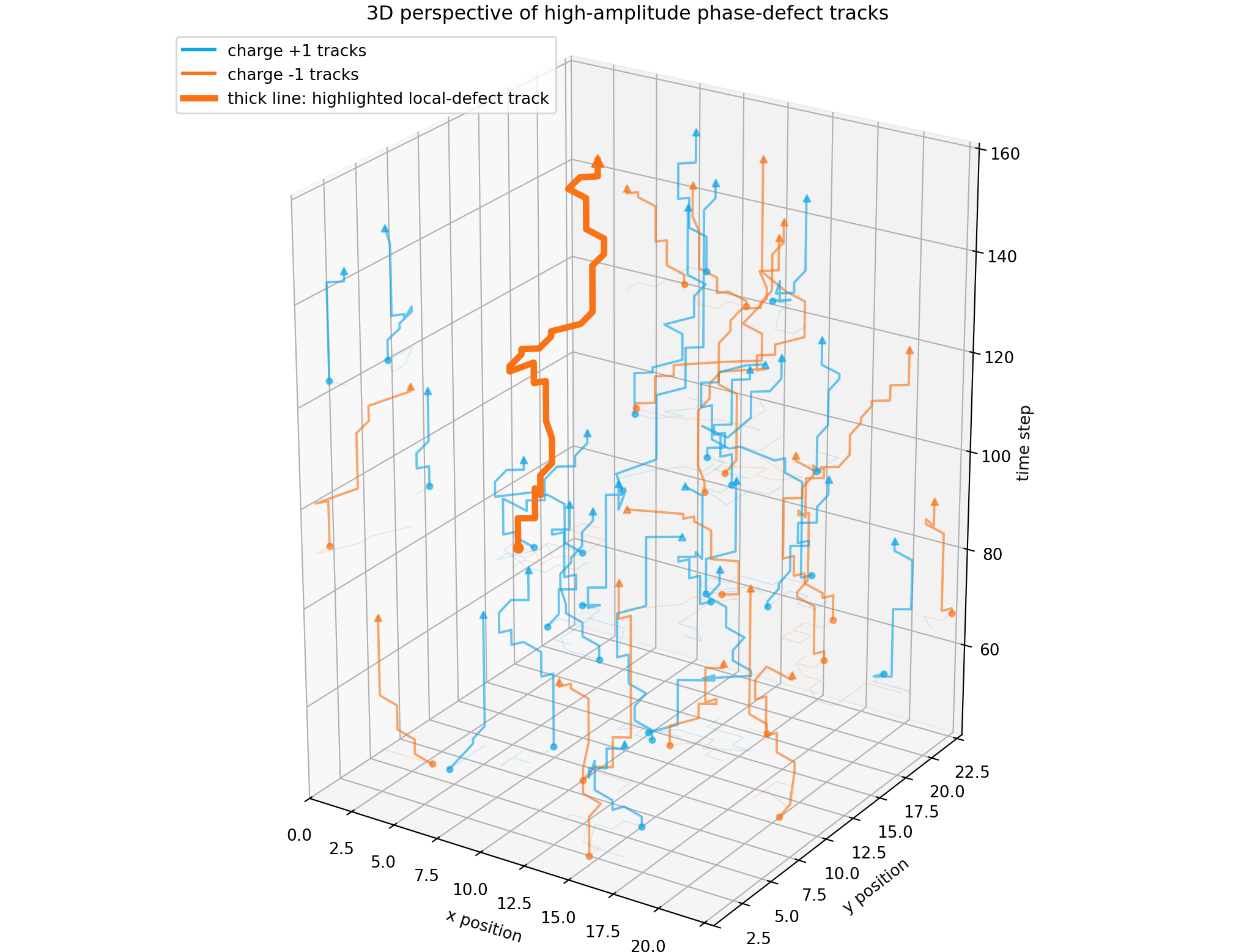}
        \caption{Three-dimensional perspective of high-amplitude phase-defect tracks in $(x,y,t)$. The thick trajectory is the same defect track used for the local zoom.}
    \end{subfigure}
    \caption{Visual audit of phase defects and vortex tracks. Single-frame phase defects are detected by local phase winding; tracks link same-charge defects across time, turning a topological singularity into a space-time trajectory.}
    \label{fig:phase_defect_track_visual}
\end{figure}

\subsection{Rate-matched spike nulls}

The previous controls change both mechanism and firing statistics. We therefore added rate-matched spike nulls that start from the exact same candidate spike tensors and preserve total spike count. A global-rate shuffle preserves the number of spikes in each frame but randomizes their spatial locations; it reduces mean vortices from 6.05 to 0.69 and long tracks from 82.5 to 8.92. Independent circular time shifts preserve each cell's spike count and spatial position but destroy simultaneity; they reduce vortices to 1.80 and long tracks to 29.25. These controls indicate that firing rate alone is not sufficient.

The fixed spatial-permutation null is more diagnostic. It preserves every spike train and all pairwise temporal synchrony but assigns cells to random spatial positions. This null retains many phase defects (5.72 vortices and 84.0 long tracks) while reducing the structure-function slope from 0.94 to 0.72. We also measured excitatory synaptic-delay transfer: for each excitatory edge, we counted whether the postsynaptic neuron spiked at the edge-specific conduction delay after a presynaptic spike, and subtracted a random-target control. The original SNN has a positive transfer excess ($1.58\times 10^{-3}$), whereas the global-rate shuffle and fixed spatial permutation are near zero ($-1.45\times 10^{-4}$ and $-1.38\times 10^{-5}$). Thus, the spatial permutation can preserve phase-defect counts while \rev{disrupting edge-aligned spike timing}. We therefore interpret turbulence-like dynamics as a joint signature of phase defects, long tracks, spatial scaling, and delayed propagation, not as a vortex-count threshold alone \rev{(Table~\ref{tab:coverage})}.

\begin{table}[t]
\centering\small
\begin{revision}
\caption{Positive contrasts (pointwise 95\% paired seed-bootstrap intervals; 20,000 resamples; 12 seeds/state). $D$: unfiltered-density $S_2$; $T$: raw-spike delay transfer; $P$: defect/track counts. Episode shifts preserve within-window counts; this is not turbulence specificity.}
\label{tab:coverage}
\begin{tabular}{lccc}
\toprule
State & Global-rate shuffle & Spatial permutation & Time shift \\
\midrule
High-rate baseline & $D,T,P$ & $D,T$ & $D,P$ \\
Low-rate episode & $D,T$ & $D,T,P$ & $T,P$ \\
\bottomrule
\end{tabular}
\end{revision}
\end{table}

\subsection{Dendritic inhibition}

We next asked whether inhibitory input location changes the turbulence-like observables. A PV-like soma-only inhibition condition reduced mean vortices to 3.40 and long tracks to 54.83. Mixed PV/SST routing recovered mean vortices to 5.95 and increased long tracks to 90.33. SST-rich dendritic inhibition produced similar vortex density and 86.0 long tracks. This supports the interpretation that perisomatic inhibition clamps firing and propagation, while dendritic inhibition reshapes the integration window in a way that better preserves spatial phase structure.

\subsection{Explicit inhibitory microcircuits}

We then replaced the lumped inhibitory population with explicit PV-like, SST-like, and VIP-like inhibitory subtypes. PV-like neurons route inhibition preferentially to the soma, SST-like neurons route inhibition preferentially to dendrites, and VIP-like neurons preferentially inhibit SST-like neurons, implementing a simplified disinhibitory gate.

This cell-type split strongly affected the spike-derived field observables. A lumped inhibitory baseline produced 6.15 mean high-amplitude vortices and 89.58 long tracks. A PV-only network nearly eliminated phase-defect dynamics, with 0.35 mean vortices and 4.08 long tracks. An SST-only network \rev{preserved measured phase structure}, with 5.93 mean vortices and 94.33 long tracks. A PV+SST network was strongly suppressed (1.07 mean vortices, 15.92 long tracks), while adding VIP partially restored phase structure (3.77 mean vortices, 62.0 long tracks). Thus, explicit inhibitory cell-type structure supports the same qualitative mechanism suggested by the routing experiment: perisomatic inhibition clamps propagation, dendritic inhibition preserves structured phase dynamics, and VIP-like disinhibition can reopen the dendritic pathway.

\subsection{Microcircuit phase diagram}

Finally, we asked whether the PV/SST/VIP result is an isolated operating point or part of a broader parameter region. We fixed the VIP fraction at 0.15 and mapped the PV fraction among non-VIP inhibitory neurons against the VIP$\to$SST disinhibition scale, using four seeds per grid point. The strongest \rev{morphology score} occurred at a low PV setting (PV axis 0.25, actual PV fraction 0.2125, actual SST fraction 0.6375) with VIP$\to$SST scale 0.5, yielding 6.37 mean vortices and 92.75 long tracks. Fully PV-dominated networks were suppressed across all VIP$\to$SST strengths, with about 0.71 mean vortices and 10.75 long tracks. Fully SST-dominated networks preserved strong vortex dynamics, but the low-PV mixed regime scored higher \rev{under the tested heuristic}.

\subsection{Short-term synaptic plasticity}

A presynaptic-type STP model showed that STP does not simply increase turbulence-like structure. Pure short-term depression reduced mean vortices to 0.41 and long tracks to 5.67. Pure facilitation increased firing rate to 163.62 Hz but did not exceed the baseline vortex density. A coarse motif with excitatory depression and inhibitory facilitation preserved long tracks (85.5) at a low firing rate (21.05 Hz), suggesting that STP can shift the network into intermittent low-rate dynamics rather than uniformly amplifying activity.

We then refined STP by projection type: $E\!\to\!E$, $E\!\to\!I$, $I\!\to\!E$, and $I\!\to\!I$. This is \rev{projection-specific} than assigning one STP rule to all excitatory or inhibitory outputs. The strongest tested motif was SST-like $I\!\to\!E$ facilitation, with 8.05 mean high-amplitude vortices, 120.42 long tracks, and 15.62 Hz mean firing rate. In contrast, a PV-like $I\!\to\!E$ depression motif nearly eliminated vortex structure, with 0.54 mean vortices and 7.67 long tracks.

Thus, the high-rate passive-dendrite baseline is not the only route to turbulence-like field structure. Projection-specific inhibitory facilitation provides a lower-rate, intermittent route in which long-lived phase-defect dynamics persist while the mean firing rate drops by more than sevenfold relative to the baseline candidate. This supports the interpretation that STP is a biophysical regulator of the operating regime rather than a cosmetic addition to the model.

\subsection{\rev{Rate-constrained low-rate STP episode}}

Because the passive-dendrite baseline has a high mean firing rate, we next made firing rate an explicit design constraint. We evaluated projection-specific STP configurations around the SST-like facilitation motif under \rev{an exploratory full-record} mean-rate window of 5--30 Hz, while requiring preserved vortices, long tracks, spatial scaling, and positive synaptic-delay transfer. Pilot calibration was followed by six-seed validation of the strongest candidate regimes.

\rev{For full 340-ms records,} the best validated candidate stayed inside the target rate window for all six pilot-validation seeds. It produced mean firing rate 12.03 Hz, population spike-count CV 1.50, 8.87 high-amplitude vortices, 129.33 tracks with lifetime at least 8 frames, structure-function slope 0.93, and positive synaptic-delay transfer $3.34\times10^{-3}$. We then fixed this parameter set and ran 12 additional seeds held out from pilot calibration. All 12 remained in the 5--30 Hz target window, with mean firing rate $12.25\pm1.21$ Hz, 8.70$\pm$0.59 high-amplitude vortices, 122.58$\pm$8.56 long vortex tracks, structure-function slope $0.94\pm0.03$, and synaptic-delay transfer $3.80\times10^{-3}\pm4.18\times10^{-4}$. This candidate therefore \rev{addresses the biological-rate objection for finite episodes}: turbulence-like field observables do not require the 115 Hz high-excitability operating point when projection-specific STP is included. \rev{Last spikes occur at $195.2\pm11.9$ ms; active-support rate is $21.31\pm2.51$ Hz (onset-to-last-spike denominator). Corrected fields yield $3.42\pm0.92$ defects, $21.08\pm9.07$ tracks and $S_2=0.768\pm0.029$. After transformation, retain [max(50 ms, stimulus offset), min(last spike+20 ms, 290 ms)]. These supersede the full-record field estimates above.}

\rev{The original full-record null analysis gave the following values.} As in the high-rate baseline, the result is not a simple vortex-count story. The global-rate shuffle preserved mean rate (12.25 Hz) and retained many phase defects (8.11 vortices, 122.75 long tracks), but its structure-function slope dropped from 0.94 to 0.73 and its excitatory synaptic-delay transfer collapsed from $3.76\times10^{-3}$ to $2.18\times10^{-4}$. A fixed spatial permutation similarly retained phase defects (7.76 vortices, 110.92 long tracks) while reducing the structure-function slope to 0.73 and eliminating synaptic-delay transfer ($-9.42\times10^{-5}$). Independent time shifts strongly reduced phase-defect dynamics (0.53 vortices, 7.75 long tracks). Thus, even in the biologically more plausible low-rate regime, smooth band-limited spike-derived fields can contain many mathematical phase singularities; the mechanistic evidence comes from their conjunction with spatial scaling and delayed propagation. \rev{Corrected-window $S_2$ rate-null contrast is 0.0337 [0.0137,0.0565]; topology reverses against this null.}

We also asked whether the $24\times24$ grid is too small to support the claim. Without retuning the low-rate STP parameters, we repeated the simulation on $16\times16$, $24\times24$, $32\times32$, and $40\times40$ grids for three seeds each. Vortex density remained nonzero across sizes (about 13.5--14.9 vortices per 1000 plaquettes), and the mean firing rate stayed near 5 Hz per neuron. \rev{With the original size-dependent fitting windows,} the structure-function slope changed with size, from 1.23 at $16\times16$ to 0.72 at $40\times40$, so this is a robustness check rather than a finite-size scaling proof. \rev{Common-lag (2--5 cells), activity-supported slopes are 0.757/0.753/0.758/0.779 (three seeds/size).} We then computed a traditional Beggs--Plenz style spike-bin avalanche diagnostic on sparse post-stimulus electrode samples. This post-stimulus diagnostic was not strong enough to claim classical criticality: each seed produced only about 8--10 traditional avalanches after sparse sampling, the size and duration tails lacked enough samples for power-law model comparison, and the branching ratio remained below 1 (0.74 at $24\times24$).

This negative result is specific to the short localized-stimulus record. To test classical spike avalanches more directly, we ran long weak-drive records with no localized stimulus: \texttt{stimulus\_amplitude\_mv}=0, homogeneous tonic drive, independent noise, 3000 ms records, and sparse 64-electrode sampling. At tonic drive 14.0 mV on $24\times24$, the mean firing rate was $6.91\pm0.18$ Hz per neuron, the pooled electrode-resample avalanche count was $1788\pm235$ per seed, the branching ratio was $0.997\pm0.001$, and the avalanche-size tail favored a power law over an exponential by $\Delta\mathrm{AIC}=736.8\pm99.6$ with exponent $\alpha=2.91\pm0.13$. Repeating the same weak-drive point on $24\times24$, $32\times32$, and $40\times40$ grids produced firing rates 6.48--6.81 Hz and branching ratios 0.994--1.003. Thus, the same SNN model class can show \rev{near-unit branching, not established criticality} under ongoing weak drive, but this is a separate assay from the localized spike-to-field turbulence experiment and still requires stronger independent-record, bootstrap, and finite-size tests. \rev{Five-family comparison favors log-normal at five of six drives and truncated power law at one.} Diagnostic plots are provided in Appendix~\ref{app:additional_diagnostics}.

\subsection{Topology control: transfer versus field re-entry}

The topology experiment is a control on interpretation rather than a claim that graph cycles alone explain brain turbulence. The previous experiments show that delayed spatial SNNs can generate long-lived phase-field dynamics, but they do not by themselves say whether these dynamics are simply direct feedforward spike transmission. We therefore performed a topology ablation while keeping the low-rate projection-STP cell and synapse parameters fixed. We compared the original recurrent exponential-distance graph, a strictly feedforward acyclic graph with only increasing $x$-direction edges, a same-degree forward-biased graph, and a same-degree loop-enhanced graph with additional reciprocal edges.

The first result is that feedforward topology favors direct pairwise spike transfer. Across six seeds, the feedforward acyclic graph had the largest delay-aligned pairwise information proxy, $3.50\times10^{-3}$ bits per presynaptic opportunity, whereas the original recurrent graph had $3.35\times10^{-4}$ and the loop-enhanced graph had $2.12\times10^{-4}$. This is consistent with a feedforward chain acting as a more direct transmitter. However, the feedforward graph eliminates closed loops: its cycle-edge fraction and random-walk return probability are both zero. In contrast, recurrent and loop-enhanced graphs have cycle-edge fraction 1.0 and nonzero return probability.

The second result is that recurrent topology supports activity re-entry after controlling firing rate. We adjusted tonic drive values for each topology and seed to match the mean firing rate near 10.5 Hz. After rate matching, the feedforward acyclic graph still had no graph cycles and had density-field re-entry probability $0.588\pm0.036$. The original recurrent graph had re-entry probability $0.736\pm0.012$, and the loop-enhanced graph had $0.757\pm0.019$. To test whether this was only an artifact of smoothing $\rho$, we repeated the measurement on raw spike trains and compared against a per-neuron spike-count matched random-time null. Raw spike re-entry exceeded the null in all topologies. The recurrent and loop-enhanced graphs had modestly higher excess over the null (0.138 and 0.134) than the feedforward acyclic graph (0.107). Thus, topology does not define turbulence-like dynamics by itself; it separates a feedforward transmission mode from a recurrent field-re-entry mode \rev{without establishing a computational advantage}.

\begin{table}[t]
    \centering
    \caption{Topology ablation separating direct pairwise spike transfer from \rev{recurrent activity re-entry}. Pairwise bits/opportunity are from the six-seed topology ablation. Cycle and re-entry metrics are from the rate-matched and raw-spike null controls.}
    \label{tab:topology_ablation}
    \small
    \begin{tabular}{lrrrr}
        \toprule
        Topology & Bits/opportunity & Cycle edges & $\rho$ re-entry & Raw spike excess \\
        \midrule
        Recurrent EDR & $3.35{\times}10^{-4}$ & 1.00 & 0.736 & 0.138 \\
        Feedforward acyclic & $3.50{\times}10^{-3}$ & 0.00 & 0.588 & 0.107 \\
        Forward-biased same-degree & $3.92{\times}10^{-4}$ & 1.00 & 0.652 & 0.118 \\
        Loop-enhanced same-degree & $2.12{\times}10^{-4}$ & 1.00 & 0.757 & 0.134 \\
        \bottomrule
    \end{tabular}
\end{table}

\section{Discussion}

The main lesson is that spike-derived turbulence-like observables are possible in a minimal spatial SNN, but only after an explicit scale transition from binary spikes to continuous density and phase fields. The strongest evidence is not any single statistic: avalanche tails and vortex counts can both be misleading, because controls can preserve power-law-favored tails or mathematical phase defects while degrading long tracks, structure-function scaling, and delayed synaptic transfer.

The avalanche results sharpen this separation. Short post-stimulus rasters are too sparse to support a Beggs--Plenz criticality claim, whereas long weak-drive records produce many spike-bin avalanches and branching ratios near one at biologically plausible rates. This is consistent with prior warnings about thresholding and subsampling \citep{touboul2010powerlaw,priesemann2009subsampling,wilting2018inferring}: localized initiation is useful for testing propagation-generated fields, while ongoing weak drive is better suited to testing spike-bin criticality.

The biological controls identify candidate gates for the field dynamics. PV-like perisomatic inhibition suppresses \rev{detected phase-defect activity}, SST-like dendritic inhibition preserves it, VIP-like disinhibition partially restores it, and projection-specific inhibitory facilitation provides a low-rate route to long-lived phase-defect tracks. The topology control adds a guardrail: feedforward graphs favor direct pairwise transfer, whereas recurrent graphs support rate-matched field re-entry, \rev{without proving enhanced information mixing}.

\rev{High-rate density lacks a 20--80 Hz peak above fitted aperiodic background; Butterworth/FFT defects/frame are 6.51/4.39 with 50-ms guards. Separately, five frozen configurations sustain low-rate organization for 2.8 s across 12 unseen seeds under different, phase-free criteria.}

\section{Limitations}

This is a simplified simulation study. The grid is two-dimensional and small, neurons are single-compartment LIF units with reduced dendritic variables, and PV/SST/VIP labels are functional motifs rather than validated cell classes. The high-rate passive-dendrite candidate should not be interpreted as typical spontaneous cortex; the low-rate STP replication shows that high rate is not necessary, but it is still \rev{a finite episode rather than a biological fit}. The phase diagram is a four-seed regional map, STP parameters are motif hypotheses, and avalanche evidence remains preliminary: continuous-field tails \rev{have conditional alternative-family and seed-bootstrap tests} \citep{clauset2009power}, while spike-bin avalanches still require bin-width, threshold, independent-record, and finite-size analyses. Finally, null models show that phase defects can survive spatial scrambling, topology ablations do not prove an empirical claim about cortical loops, and the model is not a reproduction of human fMRI or MEG measurement physics. Appendix~\ref{app:shd_readout} reports a small event-data boundary check in which the current frozen SNN does not improve task readout over raw event features.

\section{Conclusion}

We present a spike-to-field SNN framework in which delayed spatial propagation, dendritic filtering, inhibitory routing, and projection-specific STP can generate turbulence-like density and phase observables. The main result is a guarded one: long-lived phase-defect dynamics require interpretation jointly with spatial scaling, delayed synaptic transfer, firing-rate controls, and avalanche diagnostics. This provides a mechanistic testbed for brain turbulence observables, with the next step being stronger finite-size, avalanche, null-model, and empirical-recording comparisons.

\clearpage
\section*{\rev{Acknowledgments and Disclosures}}
\rev{ZeQome provided all funding, equipment and computing resources. Zequan Liu is employed by ZeQome.}

\bibliographystyle{plainnat}
\bibliography{references}

\appendix

\section{Additional diagnostic figures}
\label{app:additional_diagnostics}

This section collects supporting visual diagnostics that are useful for auditability but are not required to follow the main argument. \rev{Panels retain original windows/operators; corrected low-rate estimates supersede full-record values (Section~4.10).}

\begin{figure}[h]
    \centering
    \includegraphics[width=0.96\linewidth]{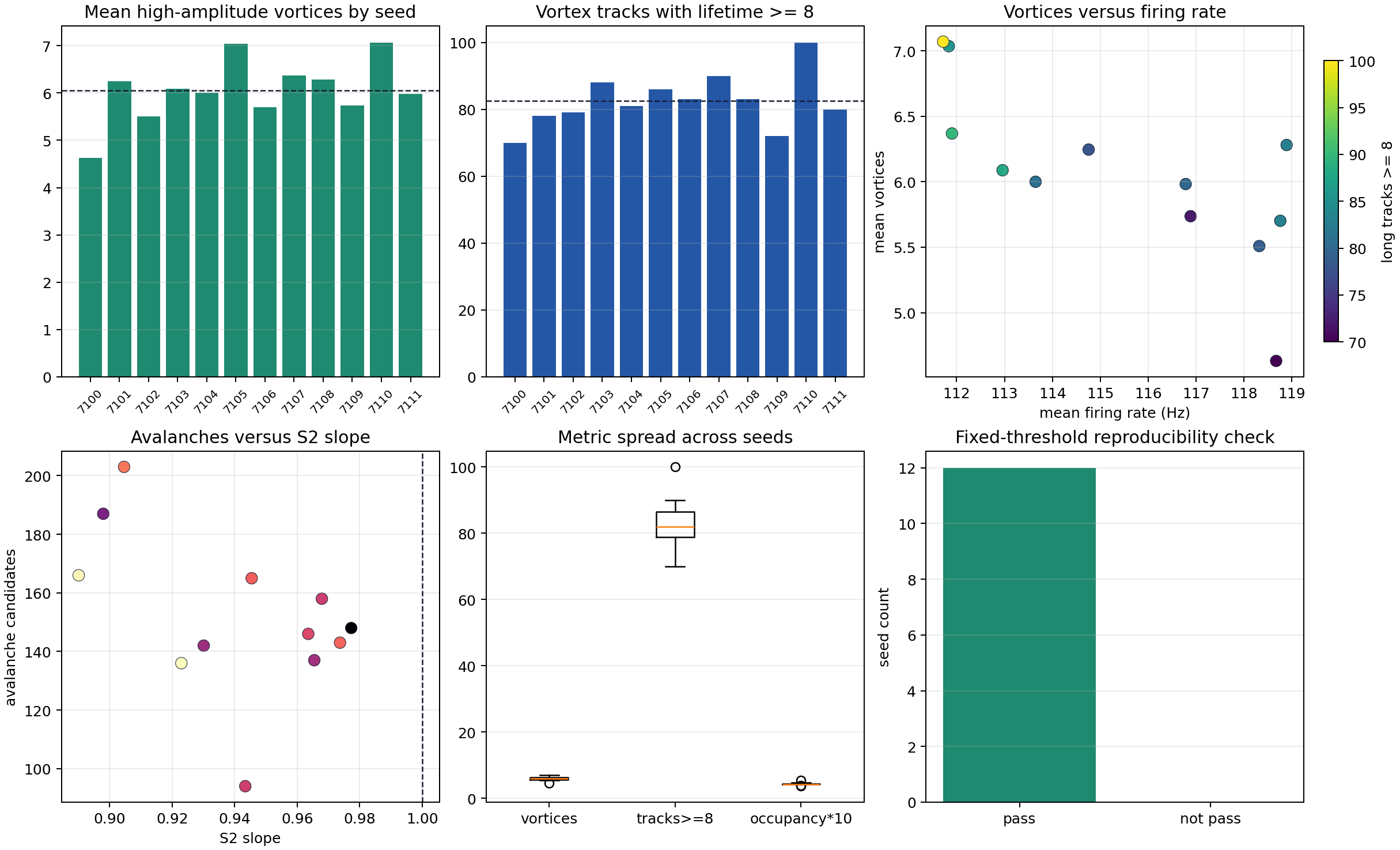}
    \caption{Multi-seed reproducibility of the candidate delayed-propagation passive-dendrite regime.}
    \label{fig:app_multiseed}
\end{figure}

\begin{figure}[h]
    \centering
    \includegraphics[width=0.96\linewidth]{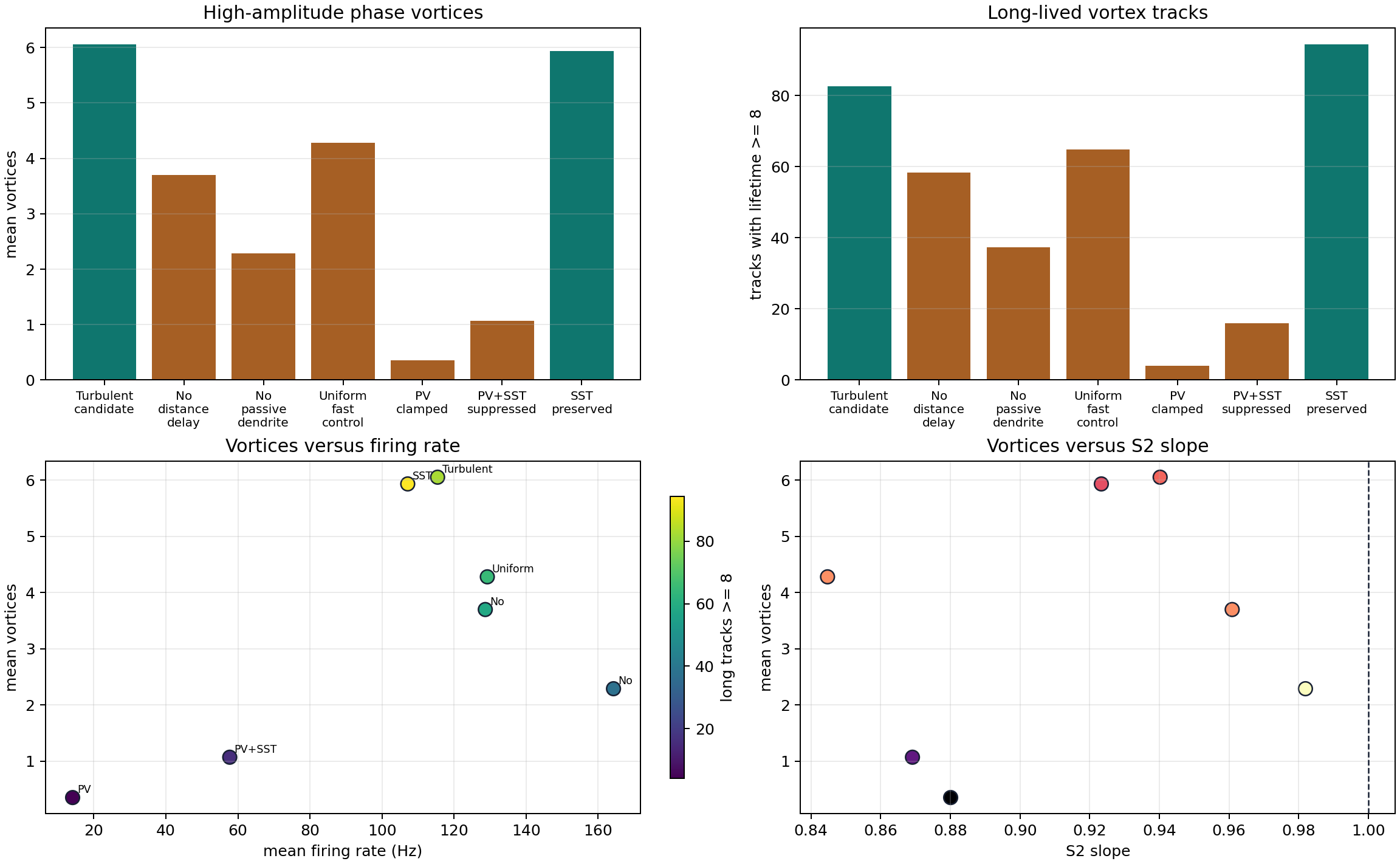}
    \caption{\rev{Baseline versus mechanism controls.}}
    \label{fig:app_turbulent_vs_non}
\end{figure}

\begin{figure}[h]
    \centering
    \includegraphics[width=0.98\linewidth]{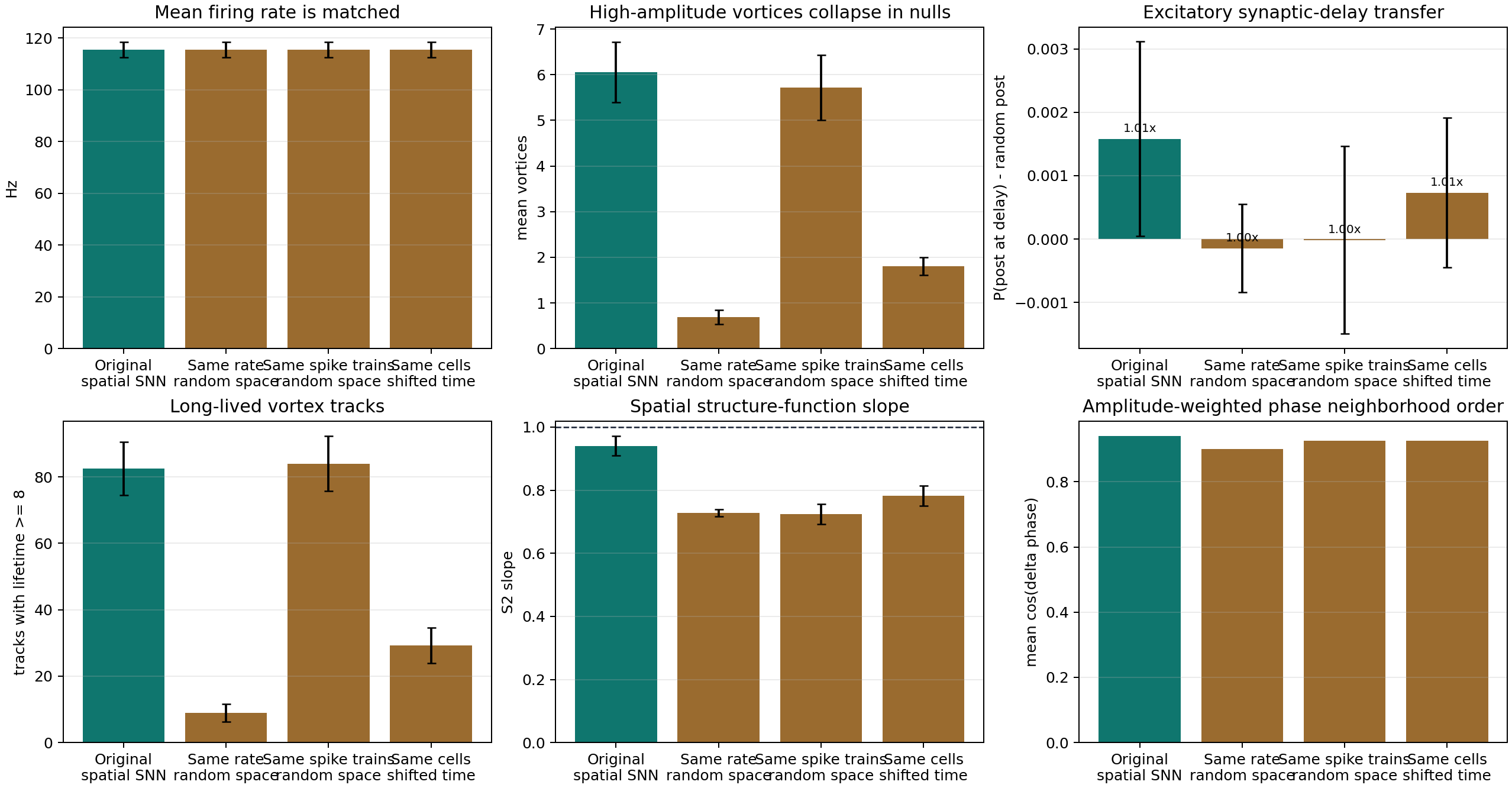}
    \caption{Rate-matched nulls constructed from the same high-rate turbulence-like SNN spike tensors. These are not separate non-turbulent SNN models: the original spike data are held fixed and then shuffled to preserve selected firing statistics while destroying selected spatiotemporal structure. The global-rate shuffle preserves the total spike count in each frame but randomizes spike locations; independent time shifts preserve each neuron's spike count and spatial identity but disrupt synchrony; the fixed spatial permutation preserves every spike train and temporal synchrony but scrambles the physical map. The comparison shows why vortex count alone is insufficient: some shuffled fields can still contain mathematical phase defects, but \rev{spatial structure and delayed transfer distinguish complementary nulls}.}
    \label{fig:app_rate_matched_nulls}
\end{figure}

\begin{figure}[h]
    \centering
    \includegraphics[width=0.96\linewidth]{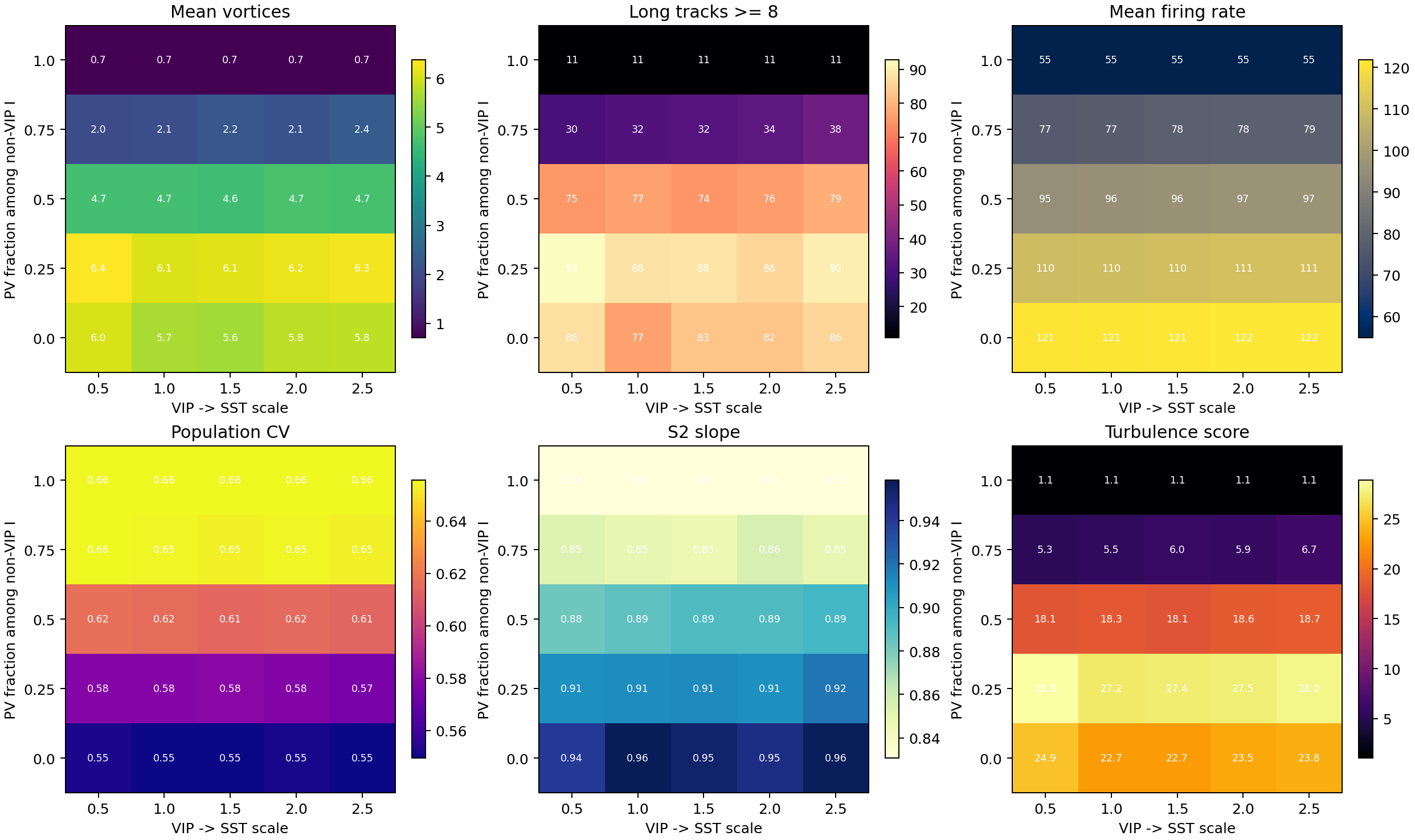}
    \caption{PV/SST/VIP microcircuit phase diagram.}
    \label{fig:app_microcircuit_phase_diagram}
\end{figure}

\begin{figure}[h]
    \centering
    \includegraphics[width=0.96\linewidth]{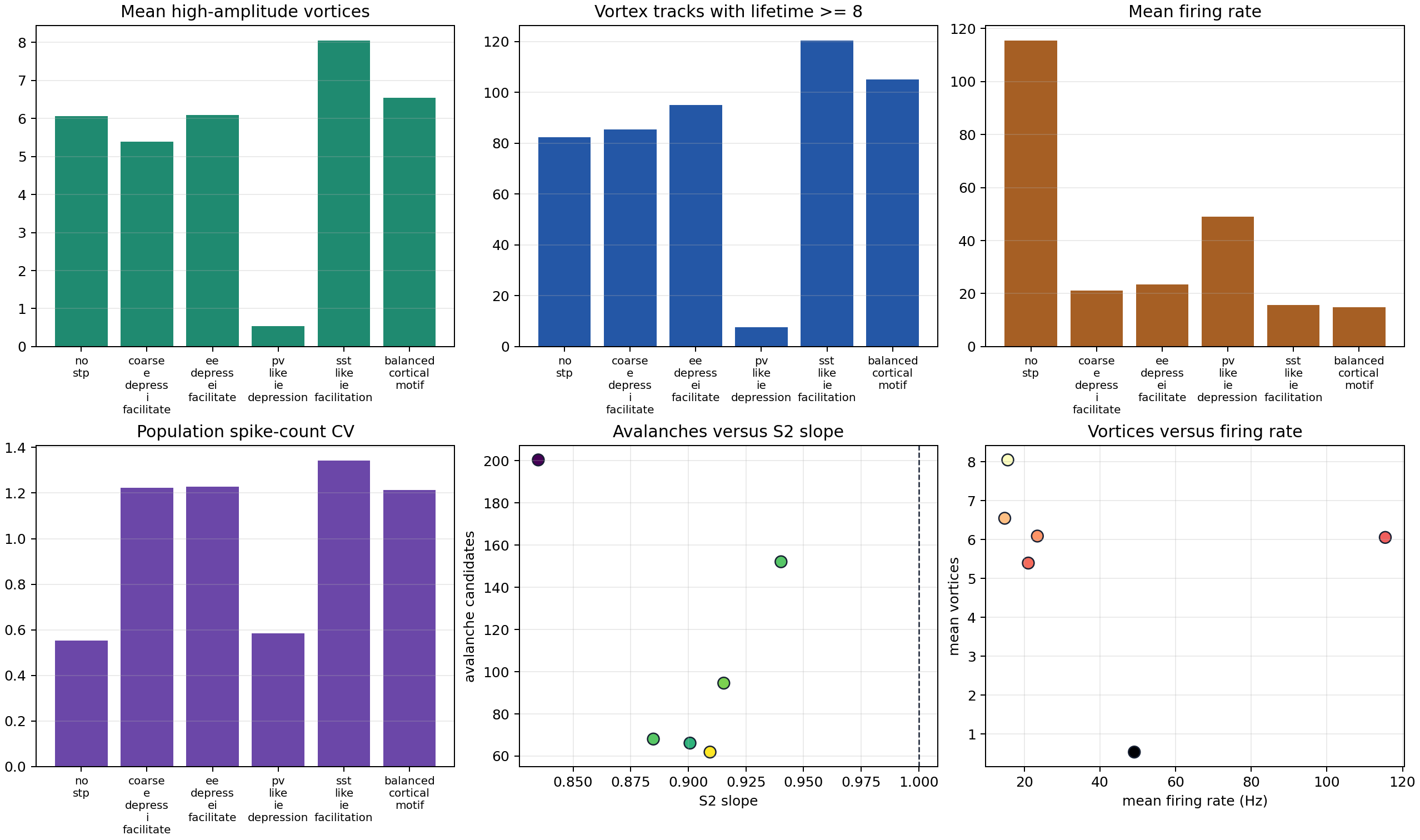}
    \caption{Projection-specific STP comparison.}
    \label{fig:app_projection_stp}
\end{figure}

\begin{figure}[h]
    \centering
    \begin{subfigure}{0.98\linewidth}
        \centering
        \includegraphics[width=\linewidth]{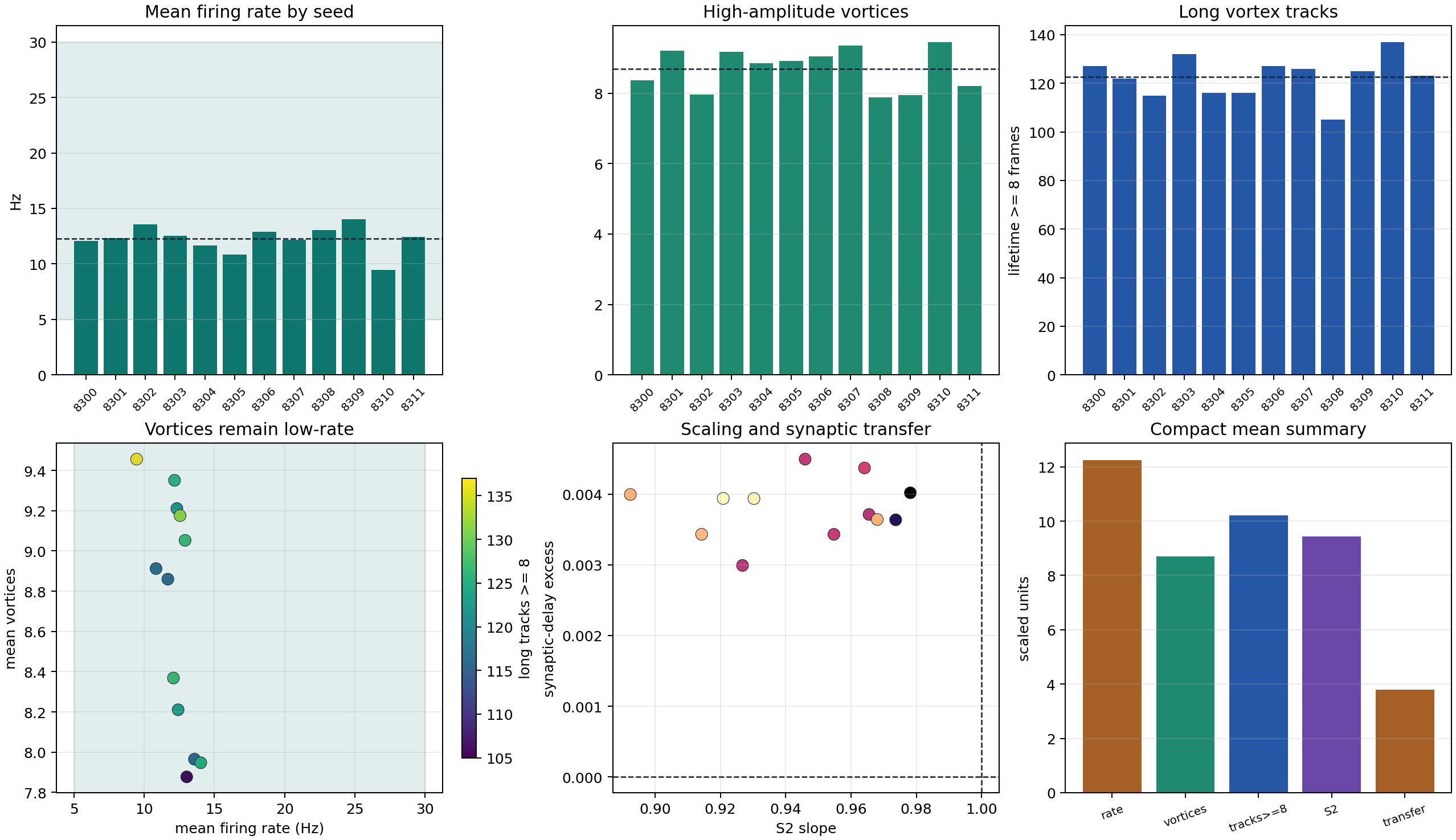}
        \caption{Twelve-seed reproducibility of the best low-rate projection-STP candidate.}
    \end{subfigure}
    \begin{subfigure}{0.98\linewidth}
        \centering
        \includegraphics[width=\linewidth]{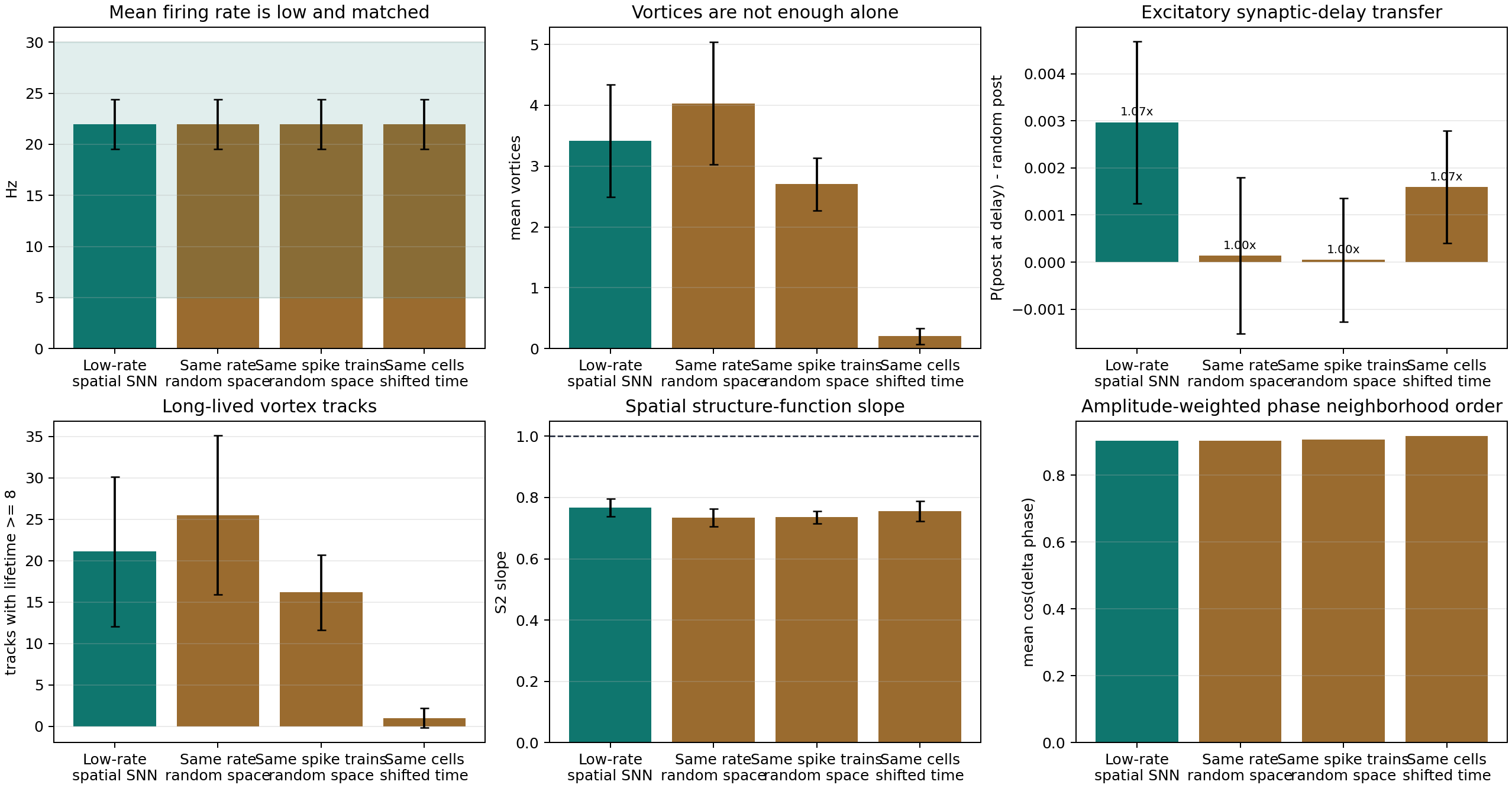}
        \caption{Rate-matched nulls for the same low-rate spike tensors.}
    \end{subfigure}
    \caption{\rev{Historical full-record episode summaries and null controls.}}
    \label{fig:app_low_rate_repro_nulls}
\end{figure}

\begin{figure}[h]
    \centering
    \includegraphics[width=0.86\linewidth]{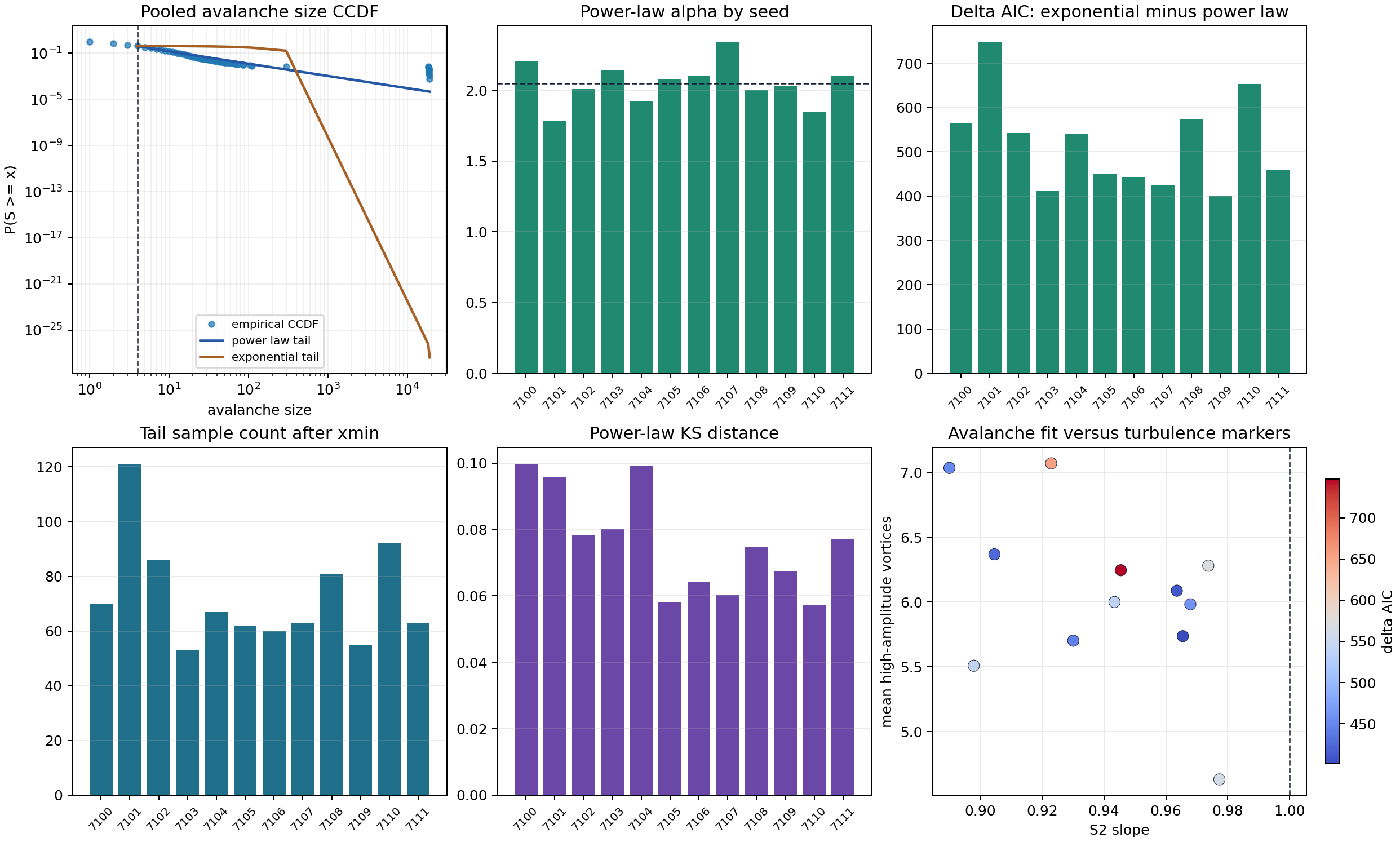}
    \caption{Continuous-field avalanche analysis. High-density connected components in $\rho(x,y,t)$ yield candidate avalanche sizes. The pooled tail is better fit by a power law than an exponential under the tested AIC comparison, but this statistic is interpreted jointly with vortex and structure-function metrics.}
    \label{fig:avalanches}
\end{figure}

\begin{figure}[h]
    \centering
    \includegraphics[width=0.96\linewidth]{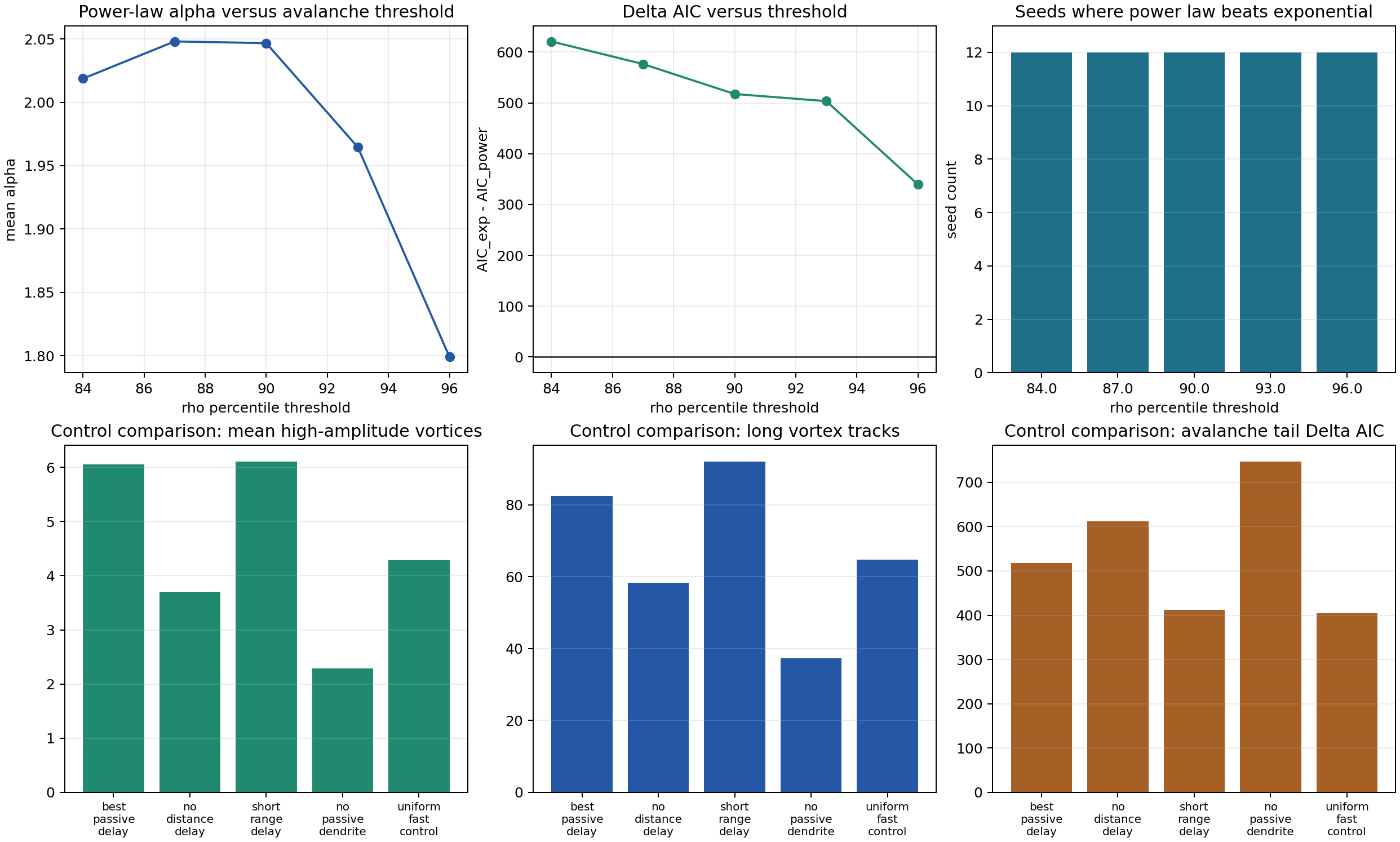}
    \caption{Threshold sensitivity and mechanistic controls. Avalanche power-law preference is robust across tested thresholds, but the phase-vortex and long-track metrics expose stronger dependence on propagation delay and passive dendritic filtering.}
    \label{fig:controls}
\end{figure}

\begin{figure}[h]
    \centering
    \includegraphics[width=0.96\linewidth]{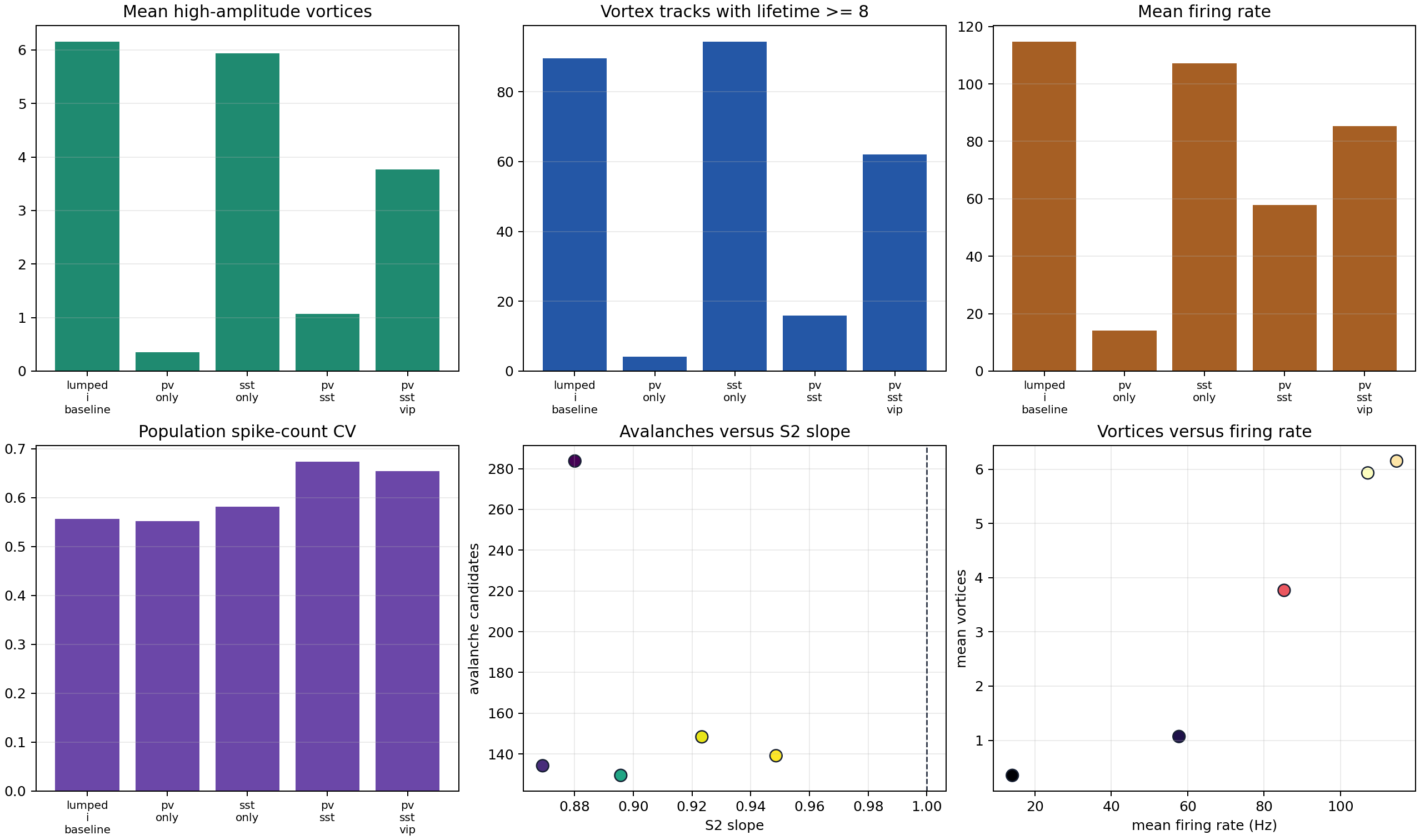}
    \caption{Explicit inhibitory microcircuit comparison. PV-only perisomatic inhibition suppresses vortex dynamics, SST-only dendritic inhibition preserves long tracks, and VIP-like disinhibition partially restores phase structure in the PV+SST setting.}
    \label{fig:microcircuit}
\end{figure}

\begin{figure}[h]
    \centering
    \includegraphics[width=0.98\linewidth]{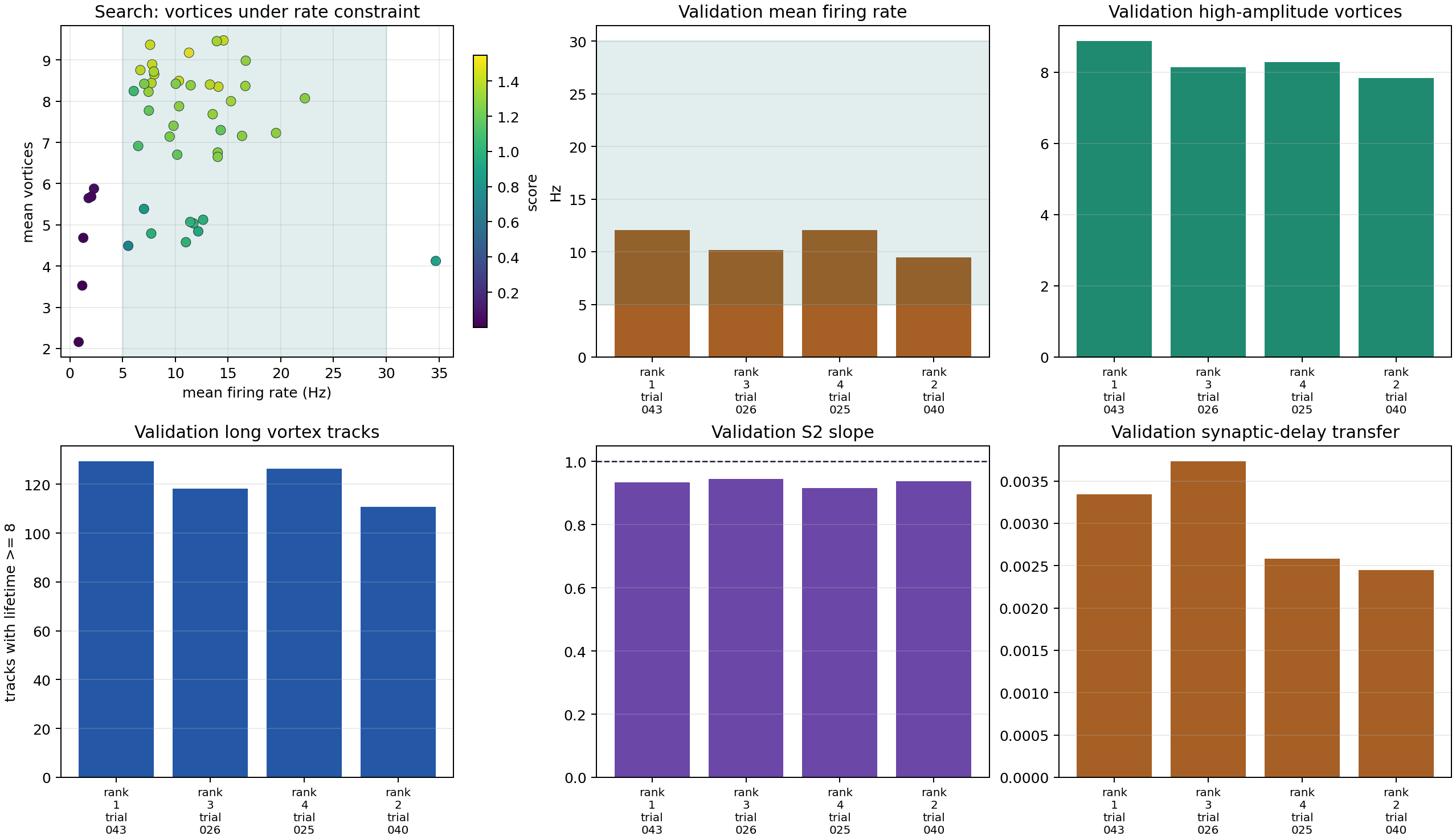}
    \caption{Rate-constrained projection-STP candidates for biologically more plausible turbulence-like regimes. The shaded firing-rate window is 5--30 Hz. The validated candidates preserve high-amplitude vortices, long tracks, spatial scaling, and positive synaptic-delay transfer \rev{at much lower full-record mean firing rates} than the passive-dendrite baseline.}
    \label{fig:rate_constrained_search}
\end{figure}

\begin{figure}[h]
    \centering
    \begin{subfigure}{0.98\linewidth}
        \centering
        \includegraphics[width=\linewidth]{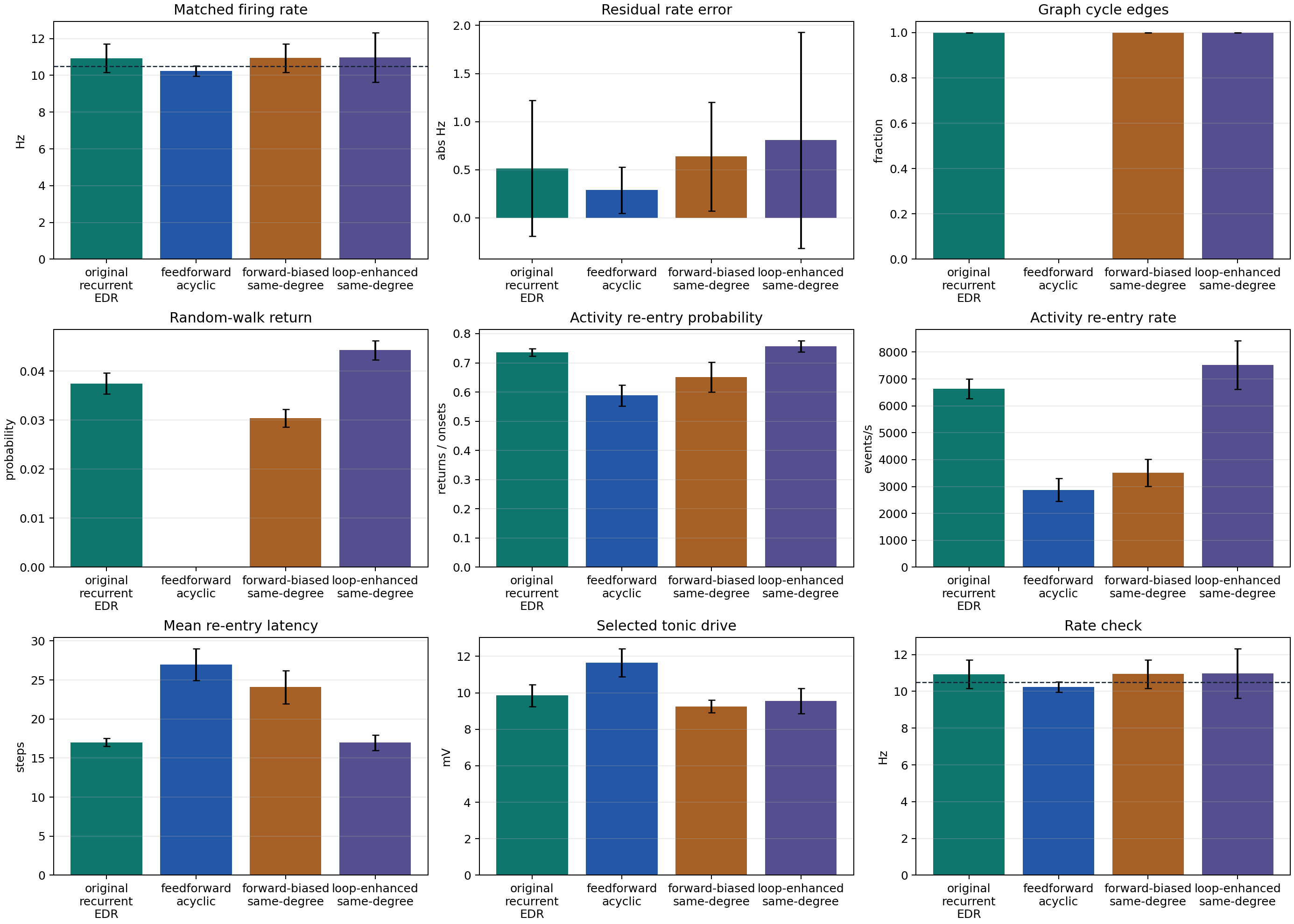}
        \caption{Rate-matched closed-loop and activity re-entry controls.}
    \end{subfigure}
    \begin{subfigure}{0.98\linewidth}
        \centering
        \includegraphics[width=\linewidth]{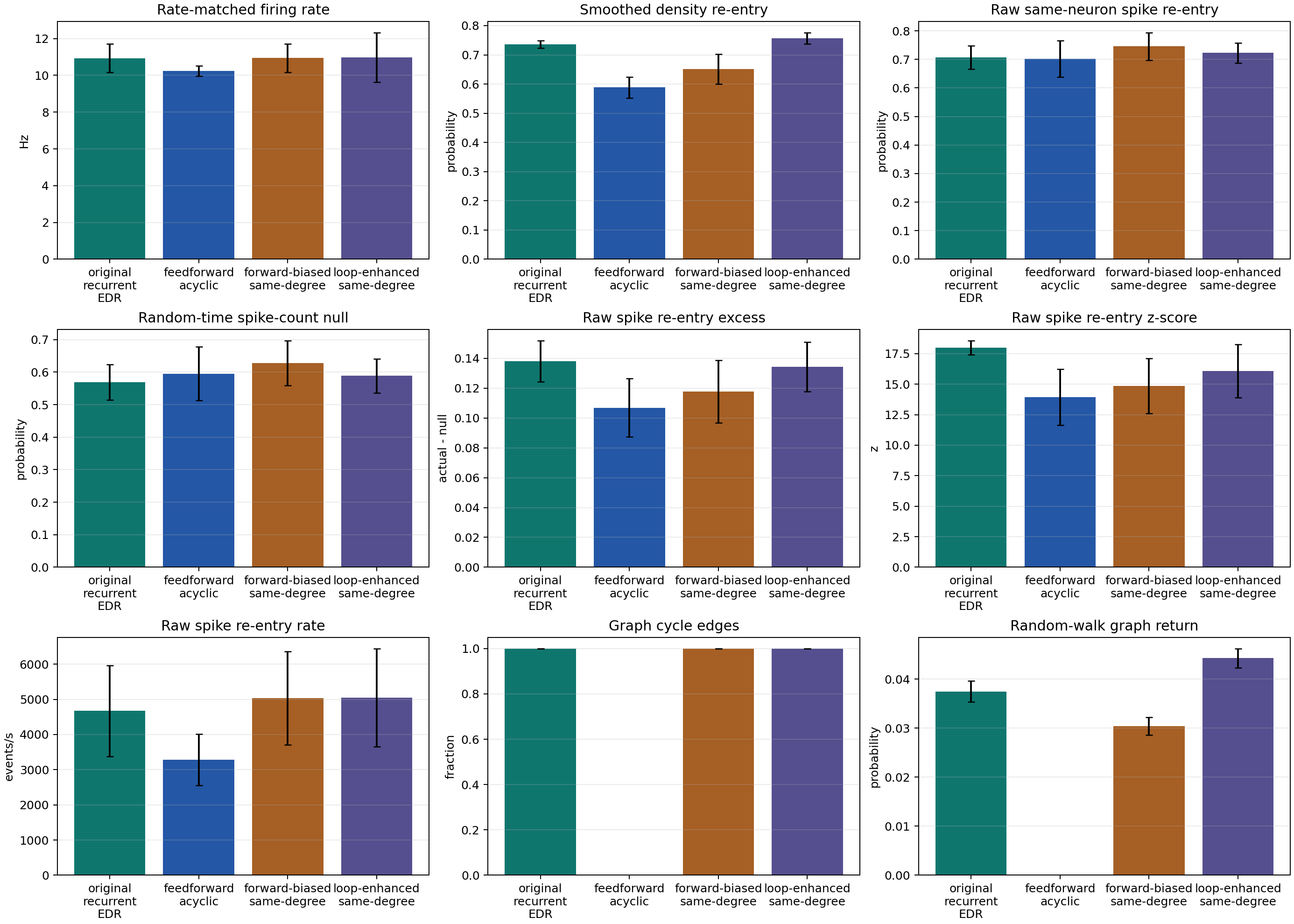}
        \caption{Raw-spike re-entry compared with per-neuron spike-count matched random-time nulls.}
    \end{subfigure}
    \caption{Topology control diagnostics. Feedforward topology improves direct delay-aligned pairwise spike transfer but removes graph cycles. Recurrent and loop-enhanced topologies preserve closed-loop structure and retain higher rate-matched activity re-entry. Raw spike re-entry also exceeds spike-count matched nulls, but the raw-spike topology effect is moderate.}
    \label{fig:cycle_reentry}
\end{figure}

\begin{figure}[h]
    \centering
    \includegraphics[width=0.98\linewidth]{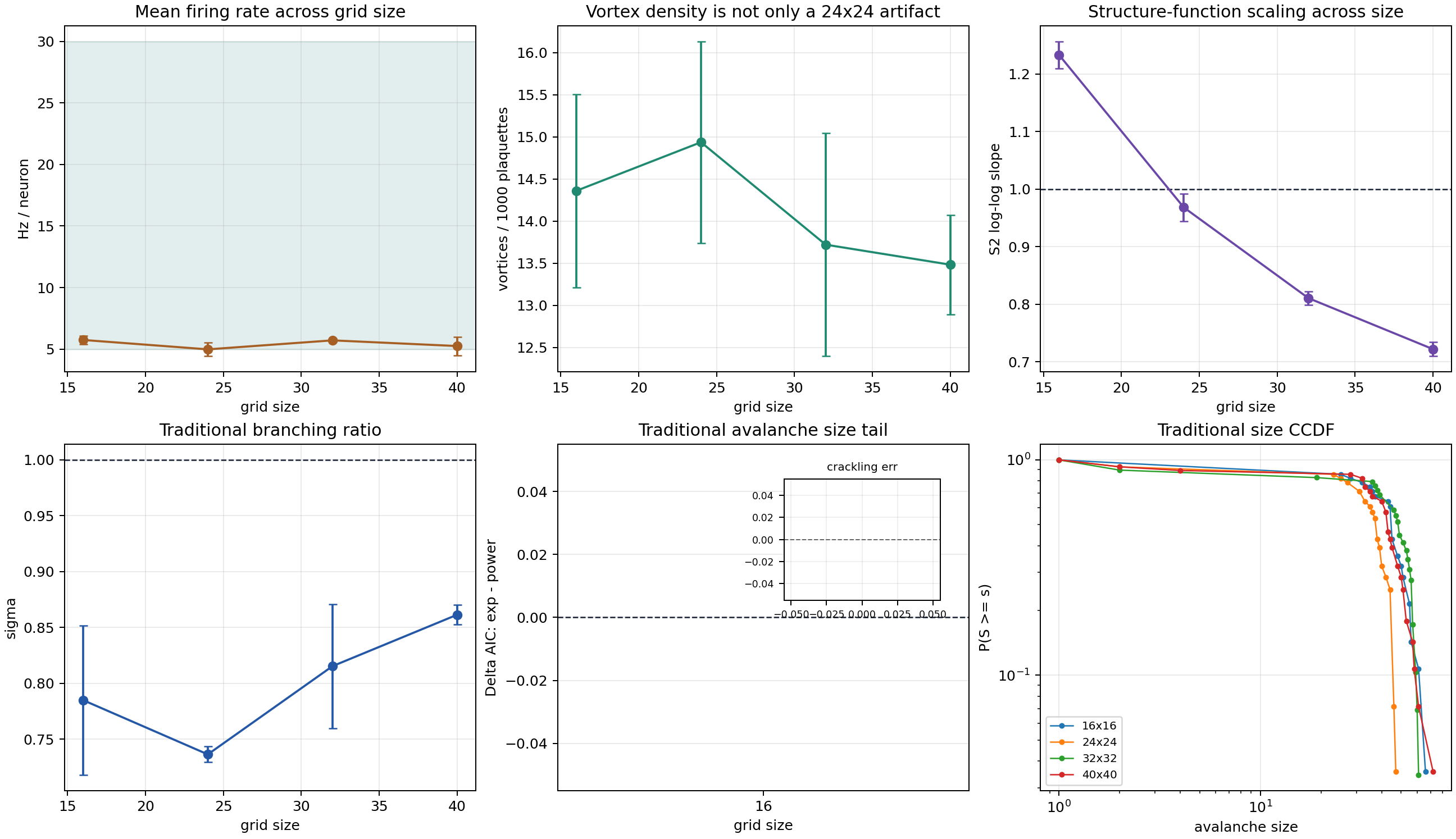}
    \caption{Traditional spike-bin avalanche and larger-grid diagnostics. The low-rate projection-STP candidate was run without retuning on $16\times16$, $24\times24$, $32\times32$, and $40\times40$ grids. Vortex density and field structure persist beyond $24\times24$, but sparse-raster traditional avalanche samples are too few for reliable size/duration power-law comparisons; branching ratios remain below 1.}
    \label{fig:traditional_avalanche_finite_size}
\end{figure}

\begin{figure}[h]
    \centering
    \includegraphics[width=0.98\linewidth]{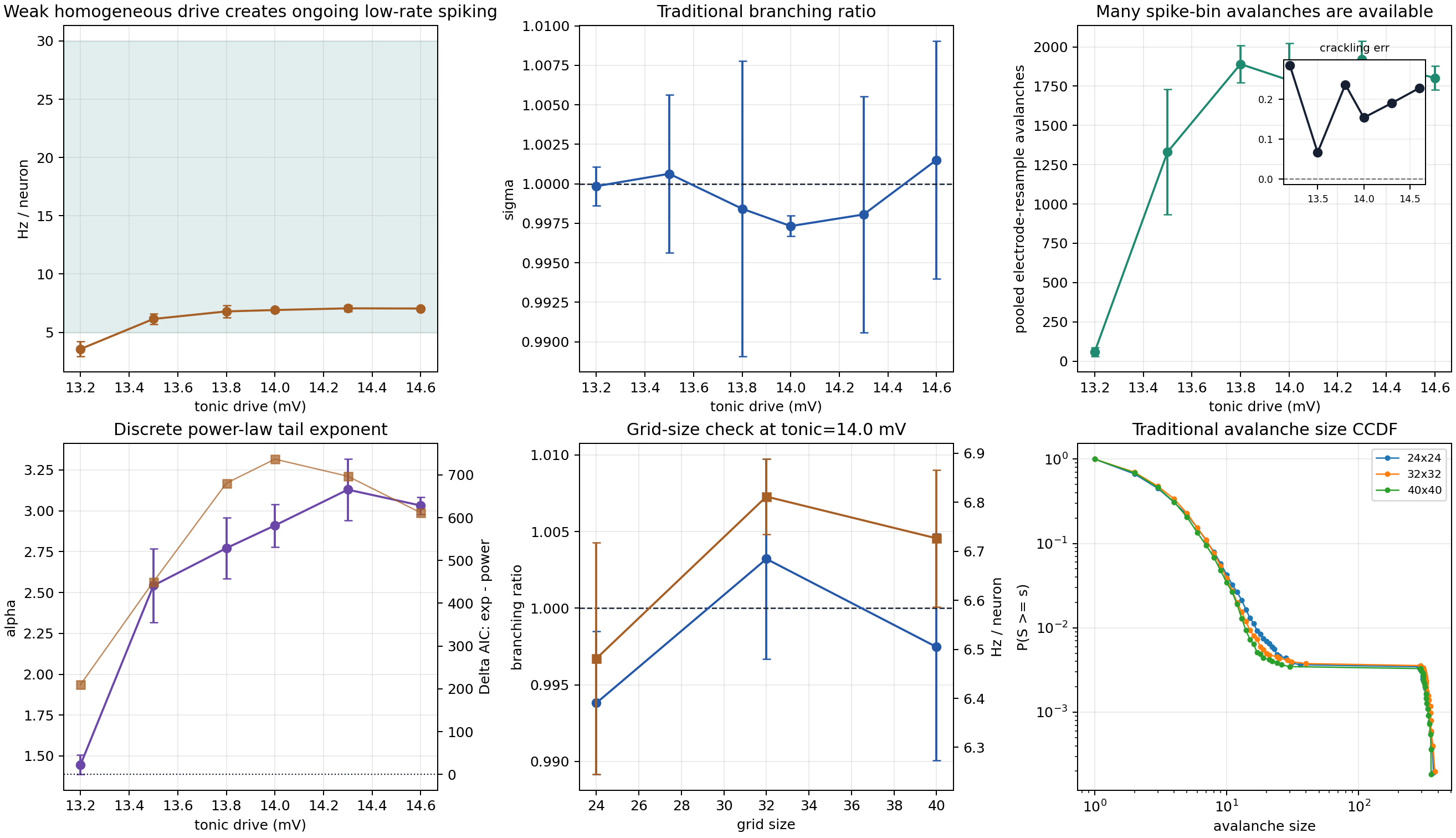}
    \caption{Traditional spike-bin avalanche diagnostics under long weak-drive spontaneous-like recording. The localized stimulus is disabled and homogeneous drive/noise maintain low-rate ongoing activity. At moderate drive, sparse-electrode spike-bin avalanches are numerous, branching ratios are near one, power-law tails are favored over exponential tails by AIC, and the effect persists from $24\times24$ to $40\times40$ grids.}
    \label{fig:spontaneous_criticality}
\end{figure}

\clearpage

\section{Real event-data readout sanity check}
\label{app:shd_readout}

The main paper studies spike-level mechanisms for turbulence-like field dynamics, not task performance. We nevertheless ran a small real event-data sanity check to test whether the current frozen SNN regimes automatically improve event-stream readout. We used the Spiking Heidelberg Digits (SHD) dataset \citep{cramer2022heidelberg}, selecting five classes with 40 training and 20 test examples. SHD is event-based audio rather than two-dimensional DVS vision; its 700 input channels were mapped onto a $28\times25$ region of the SNN input grid. This experiment is therefore a diagnostic boundary condition rather than a main benchmark.

The result is negative. With rich temporal event-bin features, a direct linear readout on raw SHD event frames achieved 0.95 test accuracy \rev{(training-CV-selected readout)} on this small subset. The frozen turbulence-like low-rate STP SNN reached \rev{0.45 linear-ridge test accuracy} using temporal state bins, the ordinary balanced E/I LIF control \rev{reached 0.45}. Thus, the present turbulence-like dynamics do not automatically improve task readout; \rev{they may mix or discard} discriminative event structure unless input projection and readout mechanisms are explicitly trained or otherwise matched to the task.

\begin{table}[h]
    \centering
    \caption{Small SHD event-data readout sanity check. These results are reported as an appendix boundary condition, not as a main contribution.}
    \label{tab:shd_appendix}
    \begin{revision}
    \begin{tabular}{lrrr}
        \toprule
        Representation / decoder & Features & CV accuracy & Test accuracy \\
        \midrule
        Raw / linear ridge & Event temporal bins & 0.625 & 0.95 \\
        Projection STP / linear ridge & State temporal bins & 0.450 & 0.45 \\
        Projection STP / RBF ridge & State temporal bins & 0.425 & 0.50 \\
        Balanced E/I / linear ridge & State temporal bins & 0.325 & 0.45 \\
        \bottomrule
    \end{tabular}
    \end{revision}
\end{table}

\begin{figure}[h]
    \centering
    \begin{subfigure}{0.98\linewidth}
        \centering
        \includegraphics[width=\linewidth]{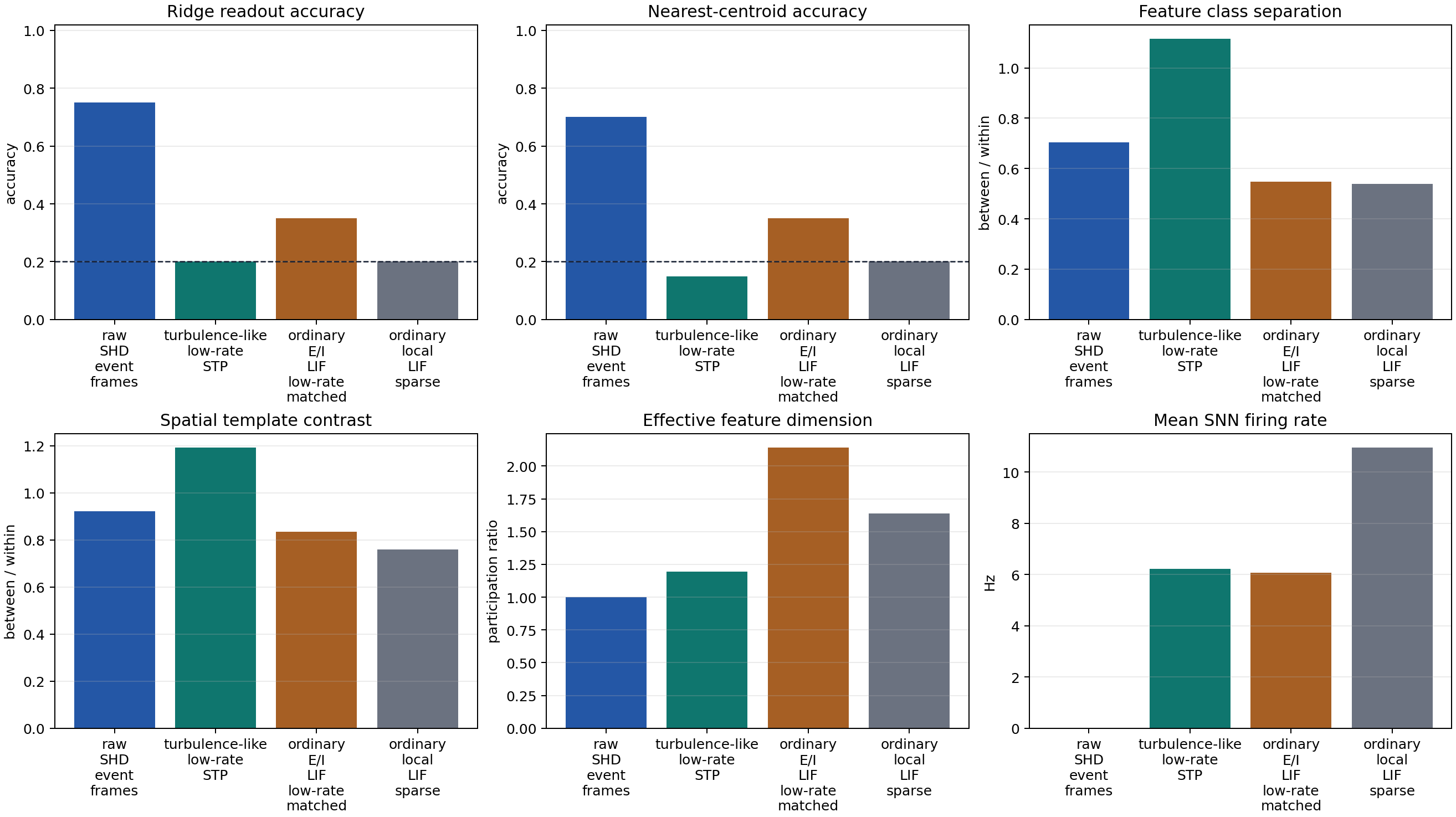}
        \caption{Information-retention diagnostics comparing raw event features and frozen SNN states.}
    \end{subfigure}
    \begin{subfigure}{0.98\linewidth}
        \centering
        \includegraphics[width=\linewidth]{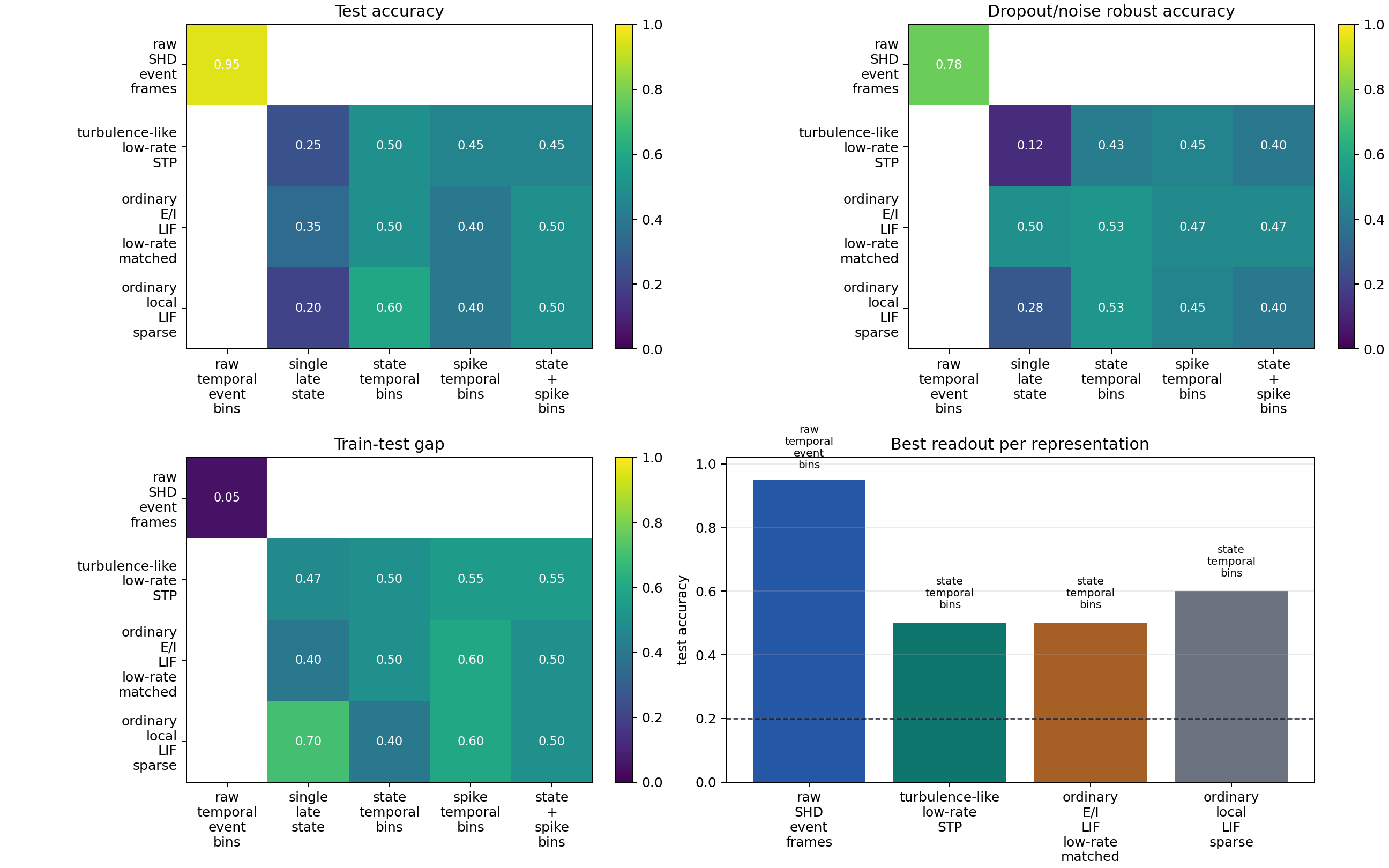}
        \caption{Readout bottleneck experiment using single-time, multi-time state, spike, and combined readout features.}
    \end{subfigure}
    \caption{Real event-data sanity check on a small SHD subset. Raw event features remain more linearly readable than the current frozen turbulence-like SNN states, reinforcing that turbulence-like dynamics should not be interpreted as automatic task-performance gains.}
    \label{fig:shd_appendix}
\end{figure}

\newpage
\input{checklist_camera_ready.tex}

\end{document}

%% file: checklist_camera_ready.tex
\section*{NeurIPS Paper Checklist}

\begin{enumerate}

\item {\bf Claims}
    \item[] Question: Do the main claims made in the abstract and introduction accurately reflect the paper's contributions and scope?
    \item[] Answer: \answerYes{}.
    \item[] Justification: The abstract and introduction state that this is a simplified SNN simulation framework and avoid claiming a direct model of biological human brain turbulence.

\item {\bf Limitations}
    \item[] Question: Does the paper discuss the limitations of the work performed by the authors?
    \item[] Answer: \answerYes{}.
    \item[] Justification: A dedicated Limitations section discusses the reduced grid, simplified neuron and dendrite models, motif-level STP assumptions, and \rev{conditional avalanche model comparisons}.

\item {\bf Theory assumptions and proofs}
    \item[] Question: For each theoretical result, does the paper provide the full set of assumptions and a complete (and correct) proof?
    \item[] Answer: \answerNA{}.
    \item[] Justification: The paper proposes an empirical simulation pipeline and does not claim formal theorems.

\item {\bf Experimental result reproducibility}
    \item[] Question: Does the paper fully disclose all the information needed to reproduce the main experimental results of the paper to the extent that it affects the main claims and/or conclusions of the paper (regardless of whether the code and data are provided or not)?
    \item[] Answer: \answerYes{}.
    \item[] Justification: The model parameters, analysis steps, and random-seed protocol are described in the paper and appendix.

\item {\bf Open access to data and code}
    \item[] Question: Does the paper provide open access to the data and code, with sufficient instructions to faithfully reproduce the main experimental results, as described in supplemental material?
    \item[] Answer: \answerNo{}.
    \item[] Justification: \rev{A review artifact exists.} The code and generated data exist in the local repository, but \rev{a complete permanent public release remains unverified}.

\item {\bf Experimental setting/details}
    \item[] Question: Does the paper specify all the training and test details (e.g., data splits, hyperparameters, how they were chosen, type of optimizer) necessary to understand the results?
    \item[] Answer: \answerYes{}.
    \item[] Justification: \rev{Simulations have no optimizer; SHD uses training cross-validation and held-out testing;} simulation parameters and candidate-regime protocol are described in the model and experiment sections.

\item {\bf Experiment statistical significance}
    \item[] Question: Does the paper report error bars suitably and correctly defined or other appropriate information about the statistical significance of the experiments?
    \item[] Answer: \rev{\answerYes{}}.
    \item[] Justification: \rev{The paper reports 12-seed means, SD and pointwise paired seed-bootstrap intervals.}

\item {\bf Experiments compute resources}
    \item[] Question: For each experiment, does the paper provide sufficient information on the computer resources (type of compute workers, memory, time of execution) needed to reproduce the experiments?
    \item[] Answer: \answerNo{}.
    \item[] Justification: The current draft does not yet report wall-clock time, CPU model, or memory usage.

\item {\bf Code of ethics}
    \item[] Question: Does the research conducted in the paper conform, in every respect, with the NeurIPS Code of Ethics \url{https://neurips.cc/public/EthicsGuidelines}?
    \item[] Answer: \answerYes{}.
    \item[] Justification: The work is a synthetic simulation study and \rev{uses no private or scraped data; SHD reuses public speech}.

\item {\bf Broader impacts}
    \item[] Question: Does the paper discuss both potential positive societal impacts and negative societal impacts of the work performed?
    \item[] Answer: \answerNo{}.
    \item[] Justification: Broader impacts have not yet been expanded \rev{here}; the current scope is foundational computational neuroscience.

\item {\bf Safeguards}
    \item[] Question: Does the paper describe safeguards that have been put in place for responsible release of data or models that have a high risk for misuse (e.g., pre-trained language models, image generators, or scraped datasets)?
    \item[] Answer: \answerNA{}.
    \item[] Justification: The paper does not release a high-risk model or scraped dataset.

\item {\bf Licenses for existing assets}
    \item[] Question: Are the creators or original owners of assets (e.g., code, data, models), used in the paper, properly credited and are the license and terms of use explicitly mentioned and properly respected?
    \item[] Answer: \answerNo{}.
    \item[] Justification: The draft cites the scientific literature, but repository and dependency license information has not yet been formalized.

\item {\bf New assets}
    \item[] Question: Are new assets introduced in the paper well documented and is the documentation provided alongside the assets?
    \item[] Answer: \answerYes{}.
    \item[] Justification: The generated simulation scripts, CSV outputs, figures, and HTML reports are documented in the local README and appendix.

\item {\bf Crowdsourcing and research with human subjects}
    \item[] Question: For crowdsourcing experiments and research with human subjects, does the paper include the full text of instructions given to participants and screenshots, if applicable, as well as details about compensation (if any)?
    \item[] Answer: \answerNA{}.
    \item[] Justification: The work does not involve \rev{new crowdsourcing or participant recruitment}.

\item {\bf Institutional review board (IRB) approvals or equivalent for research with human subjects}
    \item[] Question: Does the paper describe potential risks incurred by study participants, whether such risks were disclosed to the subjects, and whether Institutional Review Board (IRB) approvals (or an equivalent approval/review based on the requirements of your country or institution) were obtained?
    \item[] Answer: \answerNA{}.
    \item[] Justification: The work \rev{does not recruit human subjects}.

\item {\bf Declaration of LLM usage}
    \item[] Question: Does the paper describe the usage of LLMs if it is an important, original, or non-standard component of the core methods in this research? Note that if the LLM is used only for writing, editing, or formatting purposes and does \emph{not} impact the core methodology, scientific rigor, or originality of the research, declaration is not required.
    \item[] Answer: \answerNA{}.
    \item[] Justification: LLMs are not part of the \rev{simulated network}; they were used \rev{for code and manuscript assistance, subject to author verification}.

\end{enumerate}

%% file: main.bbl
\begin{thebibliography}{19}
\providecommand{\natexlab}[1]{#1}
\providecommand{\url}[1]{\texttt{#1}}
\expandafter\ifx\csname urlstyle\endcsname\relax
  \providecommand{\doi}[1]{doi: #1}\else
  \providecommand{\doi}{doi: \begingroup \urlstyle{rm}\Url}\fi

\bibitem[Beggs and Plenz(2003)]{beggs2003neuronal}
John~M. Beggs and Dietmar Plenz.
\newblock Neuronal avalanches in neocortical circuits.
\newblock \emph{Journal of Neuroscience}, 23\penalty0 (35):\penalty0
  11167--11177, 2003.
\newblock \doi{10.1523/JNEUROSCI.23-35-11167.2003}.

\bibitem[Breakspear(2017)]{breakspear2017dynamic}
Michael Breakspear.
\newblock Dynamic models of large-scale brain activity.
\newblock \emph{Nature Neuroscience}, 20:\penalty0 340--352, 2017.
\newblock \doi{10.1038/nn.4497}.

\bibitem[Clauset et~al.(2009)Clauset, Shalizi, and Newman]{clauset2009power}
Aaron Clauset, Cosma~Rohilla Shalizi, and M.~E.~J. Newman.
\newblock Power-law distributions in empirical data.
\newblock \emph{SIAM Review}, 51\penalty0 (4):\penalty0 661--703, 2009.
\newblock \doi{10.1137/070710111}.

\bibitem[Cramer et~al.(2022)Cramer, Stradmann, Schemmel, and
  Zenke]{cramer2022heidelberg}
Benjamin Cramer, Yannik Stradmann, Johannes Schemmel, and Friedemann Zenke.
\newblock The heidelberg spiking data sets for the systematic evaluation of
  spiking neural networks.
\newblock \emph{IEEE Transactions on Neural Networks and Learning Systems},
  33\penalty0 (7):\penalty0 2744--2757, 2022.
\newblock \doi{10.1109/TNNLS.2020.3044364}.

\bibitem[Deco and Kringelbach(2020)]{deco2020turbulent}
Gustavo Deco and Morten~L. Kringelbach.
\newblock Turbulent-like dynamics in the human brain.
\newblock \emph{Cell Reports}, 33\penalty0 (10):\penalty0 108471, 2020.
\newblock \doi{10.1016/j.celrep.2020.108471}.

\bibitem[Fontenele et~al.(2019)Fontenele, de~Vasconcelos, Feliciano, Aguiar,
  Soares-Cunha, Coimbra, Dalla~Porta, Ribeiro, Rodrigues, Sousa, Carelli, and
  Copelli]{fontenele2019criticality}
Antonio~J. Fontenele, Nivaldo A.~P. de~Vasconcelos, Tha{\'i}s Feliciano,
  Leandro A.~A. Aguiar, Carina Soares-Cunha, B{\'a}rbara Coimbra, Leonardo
  Dalla~Porta, Sidarta Ribeiro, Ana~Jo{\~a}o Rodrigues, Nuno Sousa, Pedro~V.
  Carelli, and Mauro Copelli.
\newblock Criticality between cortical states.
\newblock \emph{Physical Review Letters}, 122:\penalty0 208101, 2019.
\newblock \doi{10.1103/PhysRevLett.122.208101}.

\bibitem[Freeman(2009)]{freeman2009vortices}
Walter~J. Freeman.
\newblock Vortices in brain activity: Their mechanism and significance for
  perception.
\newblock \emph{Neural Networks}, 22\penalty0 (5--6):\penalty0 491--501, 2009.
\newblock \doi{10.1016/j.neunet.2009.06.050}.

\bibitem[Friedman et~al.(2012)Friedman, Ito, Brinkman, Shimono, DeVille,
  Dahmen, Beggs, and Butler]{friedman2012universal}
Nir Friedman, Shinya Ito, Braden A.~W. Brinkman, Masanori Shimono, R.~E.~Lee
  DeVille, Karin~A. Dahmen, John~M. Beggs, and Thomas~C. Butler.
\newblock Universal critical dynamics in high resolution neuronal avalanche
  data.
\newblock \emph{Physical Review Letters}, 108:\penalty0 208102, 2012.
\newblock \doi{10.1103/PhysRevLett.108.208102}.

\bibitem[Gupta et~al.(2000)Gupta, Wang, and Markram]{gupta2000organizing}
Anirudh Gupta, Yun Wang, and Henry Markram.
\newblock Organizing principles for a diversity of {GABAergic} interneurons and
  synapses in the neocortex.
\newblock \emph{Science}, 287\penalty0 (5451):\penalty0 273--278, 2000.
\newblock \doi{10.1126/science.287.5451.273}.

\bibitem[Major et~al.(2013)Major, Larkum, and Schiller]{major2013active}
Guy Major, Matthew~E. Larkum, and Jackie Schiller.
\newblock Active properties of neocortical pyramidal neuron dendrites.
\newblock \emph{Annual Review of Neuroscience}, 36:\penalty0 1--24, 2013.
\newblock \doi{10.1146/annurev-neuro-062111-150343}.

\bibitem[Markram et~al.(1998)Markram, Wang, and
  Tsodyks]{markram1998differential}
Henry Markram, Yun Wang, and Misha Tsodyks.
\newblock Differential signaling via the same axon of neocortical pyramidal
  neurons.
\newblock \emph{Proceedings of the National Academy of Sciences}, 95\penalty0
  (9):\penalty0 5323--5328, 1998.
\newblock \doi{10.1073/pnas.95.9.5323}.

\bibitem[Muller et~al.(2018)Muller, Chavane, Reynolds, and
  Sejnowski]{muller2018travelling}
Lyle Muller, Fr{\'e}d{\'e}ric Chavane, John Reynolds, and Terrence~J.
  Sejnowski.
\newblock Cortical travelling waves: Mechanisms and computational principles.
\newblock \emph{Nature Reviews Neuroscience}, 19:\penalty0 255--268, 2018.
\newblock \doi{10.1038/nrn.2018.20}.

\bibitem[Priesemann et~al.(2009)Priesemann, Munk, and
  Wibral]{priesemann2009subsampling}
Viola Priesemann, Matthias H.~J. Munk, and Michael Wibral.
\newblock Subsampling effects in neuronal avalanche distributions recorded in
  vivo.
\newblock \emph{BMC Neuroscience}, 10:\penalty0 40, 2009.
\newblock \doi{10.1186/1471-2202-10-40}.

\bibitem[Roberts et~al.(2019)Roberts, Gollo, Abeysuriya, Roberts, Mitchell,
  Woolrich, and Breakspear]{roberts2019metastable}
James~A. Roberts, Leonardo~L. Gollo, Romesh~G. Abeysuriya, Gloria Roberts,
  Philip~B. Mitchell, Mark~W. Woolrich, and Michael Breakspear.
\newblock Metastable brain waves.
\newblock \emph{Nature Communications}, 10:\penalty0 1056, 2019.
\newblock \doi{10.1038/s41467-019-08999-0}.

\bibitem[Shew and Plenz(2013)]{shew2013functional}
Woodrow~L. Shew and Dietmar Plenz.
\newblock The functional benefits of criticality in the cortex.
\newblock \emph{The Neuroscientist}, 19\penalty0 (1):\penalty0 88--100, 2013.
\newblock \doi{10.1177/1073858412445487}.

\bibitem[Spruston(2008)]{spruston2008pyramidal}
Nelson Spruston.
\newblock Pyramidal neurons: dendritic structure and synaptic integration.
\newblock \emph{Nature Reviews Neuroscience}, 9:\penalty0 206--221, 2008.
\newblock \doi{10.1038/nrn2286}.

\bibitem[Touboul and Destexhe(2010)]{touboul2010powerlaw}
Jonathan Touboul and Alain Destexhe.
\newblock Can power-law scaling and neuronal avalanches arise from stochastic
  dynamics?
\newblock \emph{PLOS ONE}, 5\penalty0 (2):\penalty0 e8982, 2010.
\newblock \doi{10.1371/journal.pone.0008982}.

\bibitem[Wilson and Cowan(1972)]{wilson1972excitatory}
Hugh~R. Wilson and Jack~D. Cowan.
\newblock Excitatory and inhibitory interactions in localized populations of
  model neurons.
\newblock \emph{Biophysical Journal}, 12\penalty0 (1):\penalty0 1--24, 1972.
\newblock \doi{10.1016/S0006-3495(72)86068-5}.

\bibitem[Wilting and Priesemann(2018)]{wilting2018inferring}
Jens Wilting and Viola Priesemann.
\newblock Inferring collective dynamical states from widely unobserved systems.
\newblock \emph{Nature Communications}, 9:\penalty0 2325, 2018.
\newblock \doi{10.1038/s41467-018-04725-4}.

\end{thebibliography}
